\input harvmac
\input epsf

%%%%%%%%%%%%%%%%%%%%%%%%%%%%%%%%%%%%%%%%%
\def\eql{~=~}

\def\al{\alpha}

\def\coeff#1#2{\relax{\textstyle {#1 \over #2}}\displaystyle}
\def\half{{1 \over 2}}

 \def\cM{{\cal M}}
\def\cN{{\cal N}} \def\cO{{\cal O}}
\def\cP{{\cal P}} \def\cQ{{\cal Q}}
\def\cR{{\cal R}} 
\def\cU{{\cal U}} \def\cV{{\cal V}}

\def\gop{\mathop{g}^{\!\circ}{}}
\def\ie{{\it i.e.}}
\def\bfone{\relax{\rm 1\kern-.35em 1}}
\def\inbar{\vrule height1.5ex width.4pt depth0pt}
\def\IC{\relax\,\hbox{$\inbar\kern-.3em{\rm C}$}}
\def\ID{\relax{\rm I\kern-.18em D}}
\def\IF{\relax{\rm I\kern-.18em F}}
\def\IH{\relax{\rm I\kern-.18em H}}
\def\II{\relax{\rm I\kern-.17em I}}
\def\IN{\relax{\rm I\kern-.18em N}}
\def\IP{\relax{\rm I\kern-.18em P}}
\def\IQ{\relax\,\hbox{$\inbar\kern-.3em{\rm Q}$}}

\def\IR{\relax{\rm I\kern-.18em R}}
\font\cmss=cmss10 \font\cmsss=cmss10 at 7pt
\def\ZZ{\relax\ifmmode\mathchoice
{\hbox{\cmss Z\kern-.4em Z}}{\hbox{\cmss Z\kern-.4em Z}}
{\lower.9pt\hbox{\cmsss Z\kern-.4em Z}}
{\lower1.2pt\hbox{\cmsss Z\kern-.4em Z}}\else{\cmss Z\kern-.4em
Z}\fi}

\def\gop{\mathop{g}^{\!\circ}{}}
\def\sech{{{\rm sech}\,}}
%
%
%%%%%%%%%%%%%%%%%%%%%%%%%%%%%%%%%%%%%%%%%
%%% Referencing
%%%%%%%%%%%%%%%%%%%%%%%%%%%%%%%%%%%%%%%%%
\def\nihil#1{{\it #1}}
\def\eprt#1{{\tt #1}}
%%%%%%%%%%%%%%%%%%%%%%%%%%%%%%%%%%%%%%%%%
\def\nup#1({Nucl.\ Phys.\ $\bf {B#1}$\ (}
\def\plt#1({Phys.\ Lett.\ $\bf  {#1B}$\ (}
\def\cmp#1({Comm.\ Math.\ Phys.\ $\bf  {#1}$\ (}
\def\prp#1({Phys.\ Rep.\ $\bf  {#1}$\ (}
\def\prl#1({Phys.\ Rev.\ Lett.\ $\bf  {#1}$\ (}
\def\prv#1({Phys.\ Rev.\ $\bf  {#1}$\ (}
\def\mpl#1({Mod.\ Phys.\ Let.\ $\bf  {A#1}$\ (}
\def\ijmp#1({Int.\ J.\ Mod.\ Phys.\ $\bf{A#1}$\ (}
\def\jag#1({Jour.\ Alg.\ Geom.\ $\bf {#1}$\ (}
\def\atmp#1({Adv.\ Theor.\ Math.\ Phys.\ $\bf {#1}$\ (}
\def\jhep#1({JHEP $\bf {#1}$\ (}

%%%%%%%%%%%%%%%%%%%%%%%%%%%%%%%%%%%%%%%%%
% References
%%%%%%%%%%%%%%%%%%%%%%%%%%%%%%%%%%%%%%%%%
%
\lref\KPW{A.\ Khavaev, K.\ Pilch and N.P.\ Warner, \nihil{New Vacua of 
Gauged  ${\cal N}=8$ Supergravity in Five Dimensions},
\plt{487} (2000) 14; \eprt{hep-th/9812035}.}
\lref\GRWlett{M.\ G\"unaydin, L.J.\ Romans and N.P.\ Warner,
\nihil{Gauged $N=8$ Supergravity in Five Dimensions,}
Phys.~Lett.~{\bf 154B} (1985) 268.}
\lref\GRW{M.\ G\"unaydin, L.J.\ Romans and N.P.\ Warner,
 \nihil{Compact and Non-Compact
Gauged Supergravity Theories in Five Dimensions,}
\nup{272} (1986) 598.}
\lref\PPvN{M.~Pernici, K.~Pilch and P. van Nieuwenhuizen,
\nihil{Gauged $N=8$, $D = 5$ Supergravity,} \nup{259} (1985) 460.}
\lref\LJR{L.J.\ Romans, \nihil{New Compactifications of Chiral $N=2$,
$d=10$ Supergravity,}  \plt{153} (1985) 392.}
\lref\RLMS{R.~G. Leigh and M.~J. Strassler, \nihil{Exactly Marginal 
Operators and
Duality in Four-Dimensional $N=1$ Supersymmetric Gauge Theory,}
\nup{447} (1995) 95; \eprt{hep-th/9503121}}
\lref\FGPWa{D.~Z. Freedman, S.~S. Gubser, K.~Pilch, and N.~P. Warner, 
\nihil{Renormalization Group Flows from Holography---Supersymmetry and a 
c-Theorem,} to appear in \atmp{3} (2000); \eprt{hep-th/9904017}.}
\lref\FGPWb{D.~Z. Freedman, S.~S. Gubser, K.~Pilch, and N.~P. Warner,
{\it Continuous Distribution of D3-branes and Gauged Supergravity,}
\jhep{0007} (2000) 38; \eprt{hep-th/9906194}. }
\lref\WestHowe{P. Howe and P. West, \nihil{The Complete 
N=2 D=10 Supergravity}, Nucl. Phys. {\bf B238} (1984) 181.}
\lref\JSIIB{J.H.\ Schwarz, \nihil{ Covariant Field Equations of 
Chiral $N=2$, $D=10$ Supergravity,} CALT-68-1016, 
\nup{226} (1983) 269.}
\lref\PvNW{P.~van~Nieuwenhuizen and N.P. Warner, \nihil{New Compactifications 
of Ten-Dimensional and Eleven-Dimensional  Supergravity on Manifolds 
which are not Direct Products} \cmp{99} (1985)  141.}
\lref\PWcrpt{K.\ Pilch and N.P.\ Warner, \nihil{A New Supersymmetric
Compactification of Chiral IIB Supergravity,} \plt{487} (2000) 22;
\eprt{hep-th/0002192}.}
\lref\PWntwo{
K. Pilch and N. P. Warner, 
\nihil{$\cN=2$ Supersymmetric RG Flows and the IIB Dilaton},
\eprt{hep-th/0004063}.}
\lref\JMNW{J.~A. Minahan and N.~P. Warner, \nihil{Quark potentials 
in the Higgs Phase of Large N Supersymmetric Yang-Mills theories,} 
\jhep{06} (1998) 005, \eprt{hep-th/9805104} .}
\lref\PolStr{J.~Polchinski and M.~J.~Strassler,
{\it The String Dual of a Confining Four-Dimensional Gauge Theory,}
\eprt{hep-th/0003136}.}
\lref\JPP{C.~V.~Johnson, A.~W.~Peet and J.~Polchinski,
{\it  Gauge Theory and the Excision of Repulson Singularities,}
Phys. Rev.\ {\bf D61} (2000) 086001, \eprt{hep-th/9911161}.}
\lref\GPPZ{L. Girardello, M. Petrini, M. Porrati and A.
Zaffaroni \nihil{The supergravity dual of N = 1 super Yang-Mills theory ,}
Nucl. Phys. {\bf B569} (2000) 451, \eprt{hep-th/9909047}.}
\lref\GPPZold{L. Girardello, M. Petrini, M. Porrati and A.
Zaffaroni \nihil{Novel local CFT and exact results on perturbations of N = 4
                  super  Yang-Mills from AdS dynamics},
JHEP {\bf 12} (1998) 022, \eprt{hep-th/9810126}.}
\lref\DistZam{J. Distler and F. Zamora, \nihil{Non-supersymmetric 
conformal field theories from stable anti-de Sitter  spaces},
Adv. Theor. Math. Phys. {\bf 2} (1999) 1405, \eprt{hep-th/9810206}.}
\lref\DistZamcor{J. Distler and F. Zamora,
\nihil{Chiral symmetry breaking in the AdS/CFT correspondence},
JHEP {\bf 05} (2000) 005, \eprt{hep-th/9911040}.}
\lref\PetZaf{M.~Petrini and A.~Zaffaroni,
\nihil{The holographic RG flow to conformal and non-conformal theory},
\eprt{hep-th/0002172}.}
\lref\CLPopecrpt{M.~Cveti\v{c}, H.~L\"u and C.N.~Pope, \nihil{Geometry of the
Embedding of Scalar Manifolds in $D=11$ and $D=10$, }
\eprt{hep-th/0002099}.}
\lref\CLPSx{
M.~Cvetic, H.~Lu, C.~N.~Pope and A.~Sadrzadeh,
\nihil{Consistency of Kaluza-Klein sphere reductions of symmetric potentials},
\eprt{hep-th/0002056}.}
\lref\CLPST{
M.~Cvetic, H.~Lu, C.~N.~Pope, A.~Sadrzadeh and T.~A.~Tran,
\nihil{Consistent SO(6) reduction of type IIB supergravity on $S^5$},
\eprt{hep-th/0003103}.}
\lref\LPTx{
H.~Lu, C.~N.~Pope and T.~A.~Tran,
\nihil{Five-dimensional N = 4, SU(2) x U(1) gauged supergravity from 
type IIB},
\eprt{hep-th/9909203}.}
\lref\CGLPx{
M.~Cvetic, S.~S.~Gubser, H.~Lu and C.~N.~Pope,
\nihil{Symmetric potentials of gauged supergravities in diverse 
dimensions and  Coulomb branch of gauge theories},
\eprt{hep-th/9909121}.}
\lref\Cetal{M.~Cvetic {\it et al.},
\nihil{Embedding AdS black holes in ten and eleven dimensions,}
Nucl.\ Phys.\  {\bf B558} (1999) 96, \eprt{hep-th/9903214}.}
\lref\BrSnfour{A.~Brandhuber and K.~Sfetsos,
\nihil{Wilson loops from multicentre and rotating branes, 
mass gaps and phase  structure in gauge theories},
\eprt{hep-th/9906201}.}
\lref\BBrSnfour{I. Bakas, A.~Brandhuber and K.~Sfetsos,
\nihil{Domain walls of gauged supergravity, M-branes, and algebraic
curves,}
\eprt{hep-th/9912132}.}
\lref\BrSntwo{A.~Brandhuber and K.~Sfetsos,
\nihil{An N = 2 gauge theory and its supergravity dual},
\eprt{hep-th/0004148}.}
\lref\BSnfour{I.~Bakas and K.~Sfetsos,
\nihil{States and curves of five-dimensional gauged supergravity},
Nucl.\ Phys.\  {\bf B573} (2000) 768, \eprt{hep-th/9909041}.}
\lref\EvPet{N.~Evans and M.~Petrini,
\nihil{AdS RG-Flow and the Super-Yang-Mills Cascade},
\eprt{hep-th/0006048}.}
\lref\AFT{
G.~Arutyunov, S.~Frolov and S.~Theisen,
\nihil{A note on gravity-scalar fluctuations in holographic RG 
flow  geometries}, \eprt{hep-th/0003116}.}
\lref\dWF{O.~DeWolfe and D.~Z.~Freedman,
\nihil{Notes on fluctuations and correlation functions in holographic
renormalization group flows,} \eprt{hep-th/0002226}.}
\lref\SGnaked{
S.~S.~Gubser,
\nihil{Curvature singularities: The good, the bad, and the naked},
\eprt{hep-th/0002160}.}
\lref\NWstrings{
N.~P.~Warner,
\nihil{Renormalization group flows from five-dimensional supergravity},
Class.\ Quant.\ Grav.\  {\bf 17}  (2000) 1287, \eprt{hep-th/9911240}.}
\lref\Porrati{M.~Porrati and A.~Starinets,
\nihil{RG fixed points in supergravity duals of 4-d field 
theory and  asymptotically AdS spaces},
Phys.\ Lett.\  {\bf B454} (1999) 77, \eprt{hep-th/9903085}.}
\lref\GPPZfine{
L.~Girardello, M.~Petrini, M.~Porrati and A.~Zaffaroni,
\nihil{Confinement and condensates without fine tuning in 
supergravity duals of gauge theories,} JHEP {\bf 9905} (1999) 026,
\eprt{hep-th/9903026}.}
\lref\KLM{A.~Karch, D.~L\"ust and A.~Miemiec,
\nihil{New N = 1 superconformal field theories and their 
supergravity  description},
Phys.\ Lett.\  {\bf B454} (1999) 265, \eprt{hep-th/9901041}.}

\lref\BalKr{V. Balasubramanian, P. Kraus and A. Lawrence,
\nihil{Bulk vs. boundary dynamics in anti-de Sitter spacetime},
Phys. Rev. {\bf D59} (1999) 046003, \eprt{hep-th/9805171}.}
\lref\CGLP{M.~Cvetic, S.~S.~Gubser, H.~Lu and C.~N.~Pope,
\nihil{Symmetric potentials of gauged supergravities in diverse dimensions
and Coulomb branch of gauge theories}, \eprt{hep-th/9909121}.}
\lref\KlWia{I.~R.~Klebanov and E.~Witten,
\nihil{AdS/CFT correspondence and symmetry breaking,}
Nucl.\ Phys.\  {\bf B556} (1999) 89, \eprt{hep-th/9905104}.}
\lref\KlWib{I.~R.~Klebanov and E.~Witten,
{Superconformal field theory on threebranes at a Calabi-Yau  singularity},
Nucl.\ Phys.\  {\bf B536} (1998) 199, \eprt{hep-th/9807080}.}
\lref\KBha{K.~Behrndt,
\nihil{Domain walls of D = 5 supergravity and fixpoints of 
N = 1 super  Yang-Mills,}
Nucl.\ Phys.\ {\bf B573} (2000) 127, \eprt{hep-th/9907070}.}
\lref\BCv{K.~Behrndt and M.~Cvetic,
\nihil{Supersymmetric domain wall world from D = 5 simple 
gauged supergravity,} Phys.\ Lett.\ {\bf B475} (2000) 253,
\eprt{hep-th/9909058}.}
\lref\SSGst{S.~S.~Gubser,
\nihil{Non-conformal examples of AdS/CFT},
Class.\ Quant.\ Grav.\  {\bf 17} (2000) 1081,
\eprt{hep-th/9910117}.}
\lref\NastVam{H. Nastase and D. Vaman, {\it On the Nonlinear KK
Reductions
on Spheres of Supergravity Theories,} \eprt{hep-th/0002028}.}
\lref\PopeCPtwo{P.~Hoxha, R.~R.~Martinez-Acosta and C.~N.~Pope,
\nihil{Kaluza-Klein consistency, Killing vectors, and Kaehler spaces,}
\eprt{hep-th/0005172}.}

\lref\PvNW{P.~van~Nieuwenhuizen and N.P. Warner, \nihil{New
Compactifications of Ten-Dimensional and Eleven-Dimensional
Supergravity on Manifolds which are not Direct Products} \cmp{99}
(1985) 141.}
\lref\MetAns{B.\ de Wit and H.\ Nicolai, \nihil{On the Relation
Between $d=4$ and $d=11$ Supergravity,} Nucl.~Phys.~{\bf B243}
(1984) 91}
\lref\deWNW{
B.\ de Wit, H.\ Nicolai and N.P.\ Warner,
\nihil{The Embedding of Gauged $N=8$ Supergravity into $d=11$
Supergravity,}
Nucl.~Phys.~{\bf B255} (1985) 29.}
\lref\WarPope{
C.~N.~Pope and N.~P.~Warner,
\nihil{
An SU(4) Invariant Compactification of D = 11 Supergravity 
on a Stretched Seven Sphere,}
Phys.\ Lett.\  {\bf B150} (1985) 352.}
\lref\CVEW{
C.~Vafa and E.~Witten,
\nihil{A Strong coupling test of S duality,}
Nucl.~Phys.~{\bf B431} (1994) 3; \eprt{hep-th/9408074}}
\lref\RDEW{
R.~Donagi and E.~Witten,
\nihil{Supersymmetric Yang-Mills Theory and Integrable
Systems,}
Nucl.~Phys.~{\bf B460} (1996) 299; \eprt{hep-th/9510101}}
\lref\dBVV{J.~de Boer, E.~Verlinde and H.~Verlinde,
\nihil{On the holographic renormalization group},
\eprt{hep-th/9912012}.}
\lref\BodCad{M.~Bodner and A.~C.~Cadavid,
\nihil{Dimensional Reduction of Type IIB Supergravity and Exceptional 
Quaternionic Manifolds}, 
Class.\ Quant.\ Grav.\  {\bf 7} (1990) 829.}
\lref\GuZaone{M.~G\"unaydin and M.~Zagermann,
\nihil{The vacua of 5d, N = 2 gauged Yang-Mills/ Einstein/ tensor 
supergravity:  Abelian case}, \eprt{hep-th/0002228}.}
\lref\GuZatwo{M.~G\"unaydin and M.~Zagermann,
\nihil{Gauging the full R-symmetry group in five-dimensional, 
$N =2$ Yang-Mills Einstein tensor supergravity}, \eprt{hep-th/0004117}.}
\lref\CerDA{A.~Ceresole and G.~Dall'Agata,
\nihil{General matter coupled $N = 2, D = 5$ gauged supergravity},
\eprt{hep-th/0004111}.}
\lref\Myers{R.C.~Myers,
\nihil{Di-electric Branes}, \jhep{9912}  (1999) 22;
\eprt{hep-th/9910053}.}
%

%%%%%%%%%%%%%%%%%%%%%%%%%%%%%%%%%%%%%%%%%

\Title{
\vbox{
\hbox{CITUSC/00-024}
\hbox{USC-00/03}
\hbox{\tt hep-th/0006066}
}}
{\vbox{\vskip -1.0cm
\centerline{\hbox{$\cN=1$ Supersymmetric Renormalization Group Flows }} 
\vskip 8 pt
\centerline{\hbox{from IIB Supergravity}}}}
\vskip -.3cm
\centerline{ Krzysztof~Pilch and Nicholas~P.~Warner }
\medskip
\centerline{{\it CIT-USC Center for Theoretical Physics}}
\centerline{{\it and}}
\centerline{{\it Department of Physics and Astronomy}}
\centerline{{\it University of Southern California}}
\centerline{{\it Los Angeles, CA 90089-0484, USA}}

\bigskip
\bigskip
We consider ${\cal N}=1$ supersymmetric renormalization group flows of
$\cN=4$ Yang-Mills theory from the perspective of ten-dimensional IIB
supergravity. We explicitly construct the complete ten-dimensional
lift of the flow in which exactly one chiral superfield becomes
massive (the LS flow). We also examine the ten-dimensional metric and
dilaton configurations for the ``super-QCD'' flow (the GPPZ flow) in
which all chiral superfields become massive. We show that the latter
flow generically gives rise to a dielectric 7-brane in the infra-red,
but the solution contains a singularity that may be interpreted as a
``duality averaged'' ring distribution of 5-branes wrapped on $S^2$.
At special values of the parameters the singularity simplifies
to a pair of $S$-dual branes with $(p,q)$ charge $(1,\pm 1)$.

\vskip .3in
\Date{\sl {June, 2000}}

%\draft

%
\parskip=4pt plus 15pt minus 1pt
\baselineskip=15pt plus 2pt minus 1pt

\newsec{Introduction}

Five-dimensional supergravities have proven to be a powerful tool in
the study of holographic RG flows of field theories on $D3$-branes.
This has been particularly well studied for the flows of $\cN = 4$
supersymmetric Yang-Mills theory under perturbations that involve
either masses or vevs for bilinear operators
\refs{\GPPZold\DistZam\KPW\KLM\GPPZfine\Porrati\FGPWa\FGPWb
\BrSnfour\KBha\CGLP\BSnfour\GPPZ\DistZamcor\SSGst
\NWstrings\dBVV\BBrSnfour\BBrSnfour
\SGnaked\PetZaf\PWntwo\BrSntwo-\EvPet}.
The corresponding five-dimensional theory is thus gauged $\cN=8$
supergravity \refs{\GRWlett\PPvN-\GRW}, but this is to be viewed as a
consistent truncation of ten-dimensional IIB supergravity
\refs{\JSIIB,\WestHowe}.  This paper will, once again, focus on such
flows of $\cN = 4$ supersymmetric Yang-Mills theory, but now from the
ten-dimensional perspective, and as we will show, this approach will
reveal some very interesting new features of these flows.

It has become increasingly evident that while the five-dimensional
theories are a valuable tool, the five-dimensional perspective is
somewhat limiting when it comes to physically interpreting the
majority of these flows.  To be more precise, almost all flows involve
running to infinite values of the supergravity potential, that is,
they are what one of our earlier collaborators dubbed {\it ``Flows to
Hades''}.  In this limit the five-dimensional supergravity metric
develops a singularity that appears superficially pathological.
However when ``lifted'' to ten dimensions, the corresponding IIB
supergravity solution  is typically rather less singular, and
may well admit a simple physical interpretation.  This softening of
the five-dimensional singularity arises partially because the ``lift''
to ten-dimensions involves multiplying the $5$-metric by a
warp-factor, and the asymptotic behaviour of the warp factor modifies
the asymptotics of the five-metric.  The simplest, but most
illustrative example of this are the $\cN = 4$ Coulomb branch flows of
\refs{\BalKr,\KlWia,\FGPWb,\BrSnfour,\CGLP}: 
In five dimensions these all generate apparently peculiar metrics with
singularities at $r=0$, whereas the corresponding ten-dimensional
metrics resolve the $r=0$ singularity into a smooth distribution of
$D3$-branes.

A second facet to the lift to ten dimensions is the IIB dilaton.  The
scalars of the five-dimensional theory are described by a coset model
${\cal S}\equiv E_{6(6)} /USp(8)$, which contains a submanifold: $
{\cal S}_0\equiv SL(6,\IR)/SO(6) \times SL(2,\IR)/SO(2)$.
Perturbatively the scalars of $SL(6,\IR)/SO(6)$ correspond to metric
perturbations on the $S^5$ of the $AdS_5 \times S^5$ compactification
of the IIB theory.  Similarly, the $SL(2,\IR)/SO(2)$ coset may be
identified with the IIB dilaton and axion at the perturbative level.
Moreover, it has been argued \refs{\FGPWb}  that this
identification remains true to all orders so long as the scalars of
the gauged $\cN=8$ supergravity are restricted to ${\cal S}_0$.  Indeed
this is strongly substantiated by the five-dimensional description of
the Coulomb branch flows.

On the other hand, it was first shown in \KPW\  that when more
general supergravity scalars are used (\ie\ ones that correspond to
fermion bilinears in the Yang-Mills theory, or corrrespond to $B_{\mu
\nu}$ fields in the IIB theory) the deformation of the $S^5$ metric is
rather more complicated.  More recently, it was also shown that when
the same supergravity scalars are non-trivial, the dilaton/axion
coset, $SL(2,\IR)_{\rm IIB}/SO(2)$, is not the same as the
$SL(2,\IR)_{5d}/SO(2)$ factor in five-dimensional coset model
$E_{6(6)} /USp(8)$.  In particular, even if the five-dimensional
scalars of $SL(2,\IR)_{5d}/SO(2)$ are set to zero it was shown in
\refs{\PetZaf,\PWntwo} that the corresponding
 ten-dimensional dilaton and axion could be highly non-trivial.

To be more explicit, it was argued in \refs{\KPW}
that the inverse metric, $g^{pq}$, on the deformed $S^5$ is given by
\eqn\metanswer{\Delta^{-{2 \over 3}}\, g^{pq} \eql 
 {1 \over a^2}\, K^{IJ\, p}\,K^{KL\, q}\, \widetilde
\cV_{IJab}\,\widetilde
\cV_{KLcd}\, \Omega^{ac}\,\Omega^{bd} \ ,}
where $\cV=(\cV^{IJab}, \cV_{I\alpha}{}^{ab})$ is the scalar matrix of
the $E_{6(6)}/USp(8)$ coset and $\widetilde\cV=(\widetilde \cV{}_{IJab},
\widetilde \cV{}^{I\alpha}{}_{ab})$ is the inverse of
$\cV$ \GRW, $K^{IJ\, p}$ are Killing vectors on $S^5$, $\Omega^{ab}$
is the $USp(8)$ symplectic form, and $\Delta = {\rm det}^{1/2}(
g_{mp}\,{\displaystyle \gop}{}^{pq})$, where ${\displaystyle
\gop}{}^{pq}$ is the inverse of the ``round'' $S^5$ metric.  The
quatity $\Delta$ can be determined by taking the determinant of both
sides of \metanswer. For more details see \refs\PWntwo.

The ten-dimensional solution is then reconstructed by taking:
\eqn\warpmetr{ds_{10}^2\eql \Omega^2 \, 
ds_{1,4}^2 ~+~ ds_5^2\,.}
where $ds^2_{1,4}$ is the metric of the $\cN=8$ supergravity
in five dimensions, $ds_5^2 = g_{mn} dy^m dy^n$ is the
deformed $S^5$ metric given by \metanswer, and 
$\Omega^2= \Delta^{-{2 \over 3}}$ is the warp factor.

In \PWntwo\ it was further argued that if $x^I$ are the cartesian 
coordinates that define the $S^5$ in $\IR^6$ (with
$\sum_I (x^I)^2 =1$), and $S$ is the IIB dilaton/axion matrix
in $SL(2,\IR)_{\rm IIB}/SO(2)$, then one has
\eqn\DilAnsatz{\Delta^{-{4 \over3}}\, (S\,S^T)^{\alpha \beta} \,=\,
{\rm const}\,\times\,\epsilon^{\alpha \gamma} \epsilon^{\beta \delta}\,
\cV_{I \gamma}{}^{ a b} \,\cV_{J \delta}{}^{ c d}
\,x^I x^J\, \Omega_{ac}\, \Omega_{bd} \, ,}
to all orders in the $\cN=8$ supergravity fields.  This is sufficient
to determine the matrix $S$ up to an $SO(2)$ gauge choice.

The argument that led to \metanswer\ showed that if consistent
truncation were true then this was necessarily the exact form of the
internal metric.  This result has since been extensively tested
\refs{\FGPWb,\CGLP,\NastVam\Cetal\CLPopecrpt\PWcrpt-%
\CLPST,\PWntwo,\BrSntwo}.   The argument that led to \DilAnsatz\ 
was similar to that for \metanswer\
but was based upon an additional (well motivated) assumption.  It has
also not been quite so well tested, but it will be implicitly tested
further by some of the results in this paper.

The distinction between the five-dimensional and ten-dimensional
$SL(2,\IR) /SO(2)$'s is a fimiliar one in field theory.  The
$SL(2,\IR)_{5d}/SO(2)$ should be viewed as the $\cN=4$ coupling at the
UV fixed point, whereas $SL(2,\IR)_{\rm IIB}/SO(2)$ should be viewed as
a running coupling of the theory on the branes.   The importance of
\DilAnsatz\ is that it gives the running coupling as an explicit
function of the UV coupling and the masses and vevs along the flow.
The derivative of \DilAnsatz\ with respect to $r$ is thus a
holographic beta function for the flow.   One should also
remember that the identification of the dilaton and axion 
as the running gauge coupling is based upon perturbation theory
about the ten-dimensional IIB theory, and thus upon
perturbations about the UV fixed point of the Yang-Mills 
theory. As in field theory, non-trivial operator mixings can 
and do  occur along RG flows, and so this running coupling may become
some other non-trivial coupling of the effective action as one
flows toward the infra-red.  Indeed, as we will see, the
flow of \FGPWa\ provides an example of this phenomenon.

The primary goal of this paper is to construct ten-dimensional
``lifts'' of two of the $\cN = 1$ supersymmetric RG flows and use then
these lifts to study the near-brane asymptotics.  The secondary
purpose of the paper is to give further support for the formula
\DilAnsatz\ by showing that it correctly predicts the dilaton
behaviour for these lifts.  

We will begin in section 2 by 
reviewing some of the the essential details of the supergravity
description of supersymmetric  RG flows, and then go on to
examine in detail the $\cN=2$  supersymmetric subsectors of 
$\cN=8$ supergravity that generate some of the possible flows
of $\cN=4$ Yang-Mills down to an $\cN=1$ theory.

In section 3 we construct the {\it complete} ten-dimensional lift of 
the ``Leigh-Strassler'' (LS) flow \refs{\RLMS,\KLM,\FGPWa}. This lift
generalizes the recent compactification of the chiral IIB
supergravity obtained in \PWcrpt\ (see, also \CLPopecrpt).  

Sections 4, 5 and 6 contain a rather involved analysis of an
$SO(3)$ invariant subsector of the $\cN=8$ supergravity in five
dimensions.   This subsector represents the truncation of $\cN=8$
supergravity down to $\cN=2$ supergravity coupled to two
hypermultiplets.  In terms of the field theory on the brane, this
sector involves breaking the $\cN=4$ Yang-Mills to $\cN=1$ with equal
masses given to each of the chiral multiplets.  A restricted version
of this was studied in detail in
\refs{\GPPZfine,\GPPZ,\DistZamcor,\PetZaf,\dWF,\AFT} from
the perspective of five-dimensional supergravity. In section
5 we examine this restriction of the the $SO(3)$ invariant sector, 
and show that it requires that the solution be $S$-dual, \ie,
self-dual with respect to $g \to 1/g$, where $g$ is the $\cN=4$
gauge coupling.  In section 6 we will then go on to construct the
ten-dimensional metric and dilaton configuration for the RG flow
(GPPZ-flow), while in section $7$ we examine the
IR asymptotics, and show how dielectric $7$-branes emerge at the
near-brane (IR) end of the  flow.  In section 8 we take a closer
look at the singularities in ten dimensions and discuss the 
relationship between our work and that of \Myers\ and \PolStr.

The reader may wish to skip from section 3 to sections 7 and
8 since the IR asymptotics can be readily understood
without pushing through the details of consistent truncations.
We have chosen to include some of the technical details
in the intervening sections, partially to facilitate 
 calculations by others working in this area, but also 
to highlight the special ``self-dual'' structure of
the flows considered in \GPPZ.  The worst of the
technical details have been relegated to an appendix.

In section 9 we construct the ten-dimensional lift of a restriction of
the GPPZ-flow to one of the fields of the LS-flow. 
While this flow is ``unphysical'' from the perspective
of the theory on the brane, it represents one of the
very few $\cN=1$ supersymmetric flows for which we have a
complete, analytically known lift.  A formal limit of
this lift reproduces the $SU(3)$ compactification of the IIB
supergravity discovered by Romans \LJR.

Finally, in  section 10 we summarize
our results, and try to draw some general threads out
of what we have learnt in using supergravity to
study RG flows holographically.

\newsec{Some $\cN=1$ supersymmetric flows}

\subsec{Supersymmetric flows in general}

As is, by now, standard we generally take the five-dimensional metric
to have the form:\foot{We use the mostly ``$-$'' convention.}
\eqn\RGFmetric{
ds^2_{1,4} = e^{2 A(r)} \eta_{\mu\nu} dx^\mu dx^\nu - dr^2 \,.}
If the supergravity scalars are canonically normalized,
with a scalar kinetic term $\half \sum_j (\del \varphi_j)^2$,
then the supersymmetric flow equations take the form:
\eqn\floweqs{ {d \varphi_j\over d r} \eql   {1 \over L}\,
{\del W \over \del \varphi_j}   \ , \qquad {d A \over d r}
\eql  - {2 \over 3 L}\,W \ .}
The supergravity potential, $\cP$, is obtained from the superpotential
via
\eqn\PandW{\cP \eql {g^2\over 8}\, \sum_j\,  \Big({\del W \over \del 
\varphi_j} \Big)^2  \,-\, {g^2\over 3}\,W^2 \,.}
In our conventions the length scale, $L$, is related to the coupling
constant, $g$, by $g=2/L$.
                                                
We will only consider flows that in the UV start in the maximally
supersymmetric vacuum with $\varphi_j = 0$. At this point both $\cP$
and $W$ have a critical point, and $W = -{3 \over 2}$.  We therefore
take $A(r) \sim r/L$ as $r \to \infty$.

As $r$ decreases there are two possibilities: either there is a ``soft
landing'' in which the flow approaches another critical point of $W$,
or $W$ decreases without bound along the flow.  If the other end of
the flow is a critical point of $W$ then one has $A(r) \sim r/\ell$ as
$r \to -\infty$, for some value of $\ell < L$, and the metric once
again approaches that of $AdS_5$.

If the flow goes to negatively infinite values of $W$, then $A'(r) \to
+\infty$, and hence $A(r) \to -\infty$, and the five-dimensional
space-time is singular.  To be more precise, the superpotential is
typically a sum of exponentials of $\varphi_j$, and one or more of the
exponentials dominate the IR limit.  One then easily solve for the
asymptotics, and one typically finds $\varphi_j \sim a_ j \log(r -
c_j)$ for some constants $a_j,c_j$.  It also turns out that $A(r) \sim
\sum b_ j \log(r - c_j)$ for some positive constants $b_j$.  As a
result, the cosmological term in \RGFmetric\ usually vanishes at
finite $r$ as some positive power: $(r-c)^{2b}$.  The power depends
upon the details, but rather little can be deduced from this behaviour
alone: One really needs the ten-dimensional solution to understand the
IR limit properly.

Our problem thus is to construct a solution to the field equations of
the chiral IIB supergravity in ten dimensions \refs{\JSIIB,\WestHowe},
\ie\ to find the metric, $g_{MN}$, the dilaton/axion field, $B$, and
the antisymmetric tensor fields, $A_{MN}$ and $F_{MNPQRS}$, expressed
in terms of the fields $A$ and $\varphi_i$ in the flow, such that the
ten dimensional equations of motion%
\foot{We refer the reader to \JSIIB\ and to our recent
paper \PWntwo\ for the explicit form of those equations.}
become equivalent to the flow
equations \floweqs.

\subsec{Supersymmetric flows in particular: Truncations}

A standard process by which one reduces the number of
supergravity scalar
fields to a more manageable subset is to impose invariance
under a carefully chosen discrete or continuous symmetry of
the action. The idea is that since the symmetry is an
invariance of the action, any expansion of the action will
be at least quadratic in non-singlet fields, and so it
is consistent to set all non-singlet fields to zero, and 
solve the equations on the space of singlets alone.
In this paper we will employ two such trunactions to arrive
at distinct $\cN=2$ supergravities, coupled to
matter and vector multiplets, and embedded in the $\cN=8$
theory.  Such $\cN=2$ supergravities are certainly not
new: there is a well established technology for constructing
broad classes of such theories (see, e.g.,
\refs{\GuZaone,\CerDA,\GuZatwo}).  
The significance of the $\cN=2$ supergravities considered here 
is that they are dual to distinct $\cN=1$
Yang-Mills theories arising from massive flows of the
$\cN=4$ theory.    

The natural way to accomplish this is to use the $SO(6) \times
SL(2,\IR)$ symmetry, and under the $SO(6)$ the gravitini
transform as $\bf 4$ and $\bf \bar 4$. To get an $\cN=2$
supergravity we use symmetries for which the $\bf 4$ has
only one singlet. An obvious candidate is to take $SU(3)
\subset SO(6)$, under which $\bf 4 \to 3 \oplus 1$.
Imposing $SU(3)$ invariance is, however, far too restrictive
and leaves only one nontrivial scalar field. (This restricted
case will, in fact, be discussed in detail in section 9.) 
Instead we pass
to subgroups of this $SU(3)$.  In the first instance
we will impose invariance under $SU(2) \times U(1) \subset
SU(3)$, and in the second we impose invariance under
$SO(3) \subset SU(3) \subset SO(6)$, where the $SO(3)$
is the real subgroup of $SU(3)$.

Considering the entire spectrum of the gauged
$\cN=8$ supergravity, the space of $SU(2) \times U(1)$ 
singlets consists of the following: a graviton,  two gravitini,  
two vector fields, no tensor gauge fields, four spinors, 
and five scalars. These make up the 
$\cN=2$ supergravity multiplet coupled to one
vector multiplet and one hypermultiplet.
A more careful examination of the group theory shows
that the scalar manifold is:
\eqn\scalmanLS{
{\cal S}_{LS} \eql {SU(2,1)\over SU(2) \times U(1) } 
 \times SO(1,1)\,.}
An $SU(1,1)$ subgroup of $SU(2,1)$ represents the dilaton/axion 
coset, while the other two non-compact generators of
$SU(2,1)$ are the supergravity dual
of a (complex) Yang-Mills fermion mass.  The $SO(1,1)$ represents a 
diagonal element of the $SL(6,\IR) \subset E_{6(6)}$ and is dual to a
Yang-Mills scalar mass.  The dilaton/axion scalars will remain 
fixed in the flows considered here, and the corresponding
reduction of the scalar manifold can be done by imposing 
invariance under another
$U(1)$ so that the coset becomes ${SU(1,1)\over U(1) } 
\times SO(1,1)$.  As in \FGPWa, 
this can be parametrized by two real scalars, $\chi$ and $\alpha$,
along with the  $U(1)$ symmetry of the denominator.
 
The $SO(3)$ invariant subsector of the $\cN=8$ theory
consists of: a graviton,  two gravitini,  one vector field,
no tensor gauge fields, four spinors, and eight scalars.
The result is thus
$\cN=2$ supergravity coupled to two hypermultiplets.
The scalar manifold of this $\cN=2$ theory is now 
the quaternionic  manifold:
\eqn\scalmanQCD{
{\cal S}_{QCD} \eql {G_{2(2)}\over SU(2)\times SU(2)}\,.}
We will discuss the parametrization of this extensively in
sections 4 and 5.

\newsec{The LS-flow}

The $SU(2) \times U(1)$ invariant sector of the 
supergravity is dual to $\cN=4$ Yang-Mills perturbed by
the bilinear operators with the same invariance.
Specifically $\chi$ and $\alpha$ are, respectively, 
dual to a single fermion mass and the mass of the
scalar\foot{The $AdS_5$-normalizable modes of the scalar $\alpha$ can
additionally be interpreted in terms of vevs of scalars in the two
massless chiral multiplets.} in the same chiral multiplet.  
The $SU(2)$ symmetry is a global symmetry of the two
remaining massless chiral multiplets, while the $U(1)$ essentially
gives rise to the $\cN=1$ $R$-symmetry.  The $\cN=1$
flows in this sector thus include the flow considered by
Leigh and Strassler in \RLMS.
 
\subsec{The five-dimensional flow}

The field $\chi$ is canonically normalized, whereas $\alpha$ is 
not:  The kinetic term is $-\half (\del \chi)^2 -3(\del \alpha)^2$.
The superpotential is:
\eqn\Wreduced{W~=~ {1 \over 4 \rho^2}~ \Big[\cosh(2 \chi)~
( \rho^{6}~-~ 2)~ - ( 3\rho^{6} ~+~ 2 ) \Big] \ .}
where $\rho =\exp(\alpha)$.  The resulting field equations are:
\eqn\fieldeqs{\eqalign{
{d\chi\over d r}&\eql {g\over 2}\,{\partial W\over \partial\chi}\eql
{g\over 4} \,{(\rho^6-2)\,\sinh(2\chi)\over
\rho^2}\,,\cr
{d\rho\over d r}&\eql {g\over 12}\,\rho^2\,{\partial W\over \partial \rho}
\eql  {g\over 12}\, {\rho^6\,(\cosh(2\chi)-3)+2\cosh^2\chi
        \over \rho}\,.\cr} }

The superpotential \Wreduced, and the corresponding potential \PandW\
have an $\cN=2$ supersymmetric critical point for $\chi =
\coeff{1}{2}\log(3)$ and $\alpha = \coeff{1}{6}\log(2)$  \KPW.
As shown in \refs{\KLM,\FGPWa}, this critical point is the dual of the
Leigh-Strassler conformal fixed point of \RLMS.  The compactification
of the chiral IIB supergravity corresponding to the critical point has
recently been obtained in \PWcrpt.  The flow itself can be obtained by
solving \fieldeqs\ with the proper initial conditions in the UV
\FGPWa.  

One of the difficulties in studying this flow is that an explicit
solution to \fieldeqs\ in a closed form is not known. Formally, one
can derive a series solution for the trajectory, $\rho=\rho(\chi)$,
which is of the form
\eqn\sersol{\eqalign{
\rho(\chi)&\eql \sum_{m=0}^\infty \sum_{n=2m}^\infty
a_{mn}\,(\log\chi)^{n}\,\chi^{2m}\cr
& \qquad \eql 1+(\gamma-\coeff23\log\chi)\chi^2
\cr&\qquad\qquad +
\left(\coeff{16 + 42 \gamma  + 171 \gamma^2}{18}-
 \coeff{2(7+57\gamma)}{9}\log\chi+\coeff{38}{9}(\log\chi)^2\right) 
\chi^4+\ldots\cr}
}
where $\gamma$ is an integration constant parametrizing the
trajectory. For futher numerical analysis we refer the reader to
\refs{\FGPWa,\SGnaked}.

There exists also a closely related $\cN=2$ flow which can be lifted
to a solution of the chiral IIB supergravity in ten dimensions
\PWntwo\ for which the analogue of the series solution \sersol\ {\it
can} be summed in terms of elementary functions
\refs{\PWntwo,\BrSntwo}.  In the following we will use the general
structure of those two solutions to obtain a ten dimensional lift of the
present flow.

\subsec{The lift to ten dimensions}

As discussed in the introduction, both the ten-dimensional metric and
the dilaton/axion field are given by the consistent truncation ansatz
\warpmetr\ and \DilAnsatz, respectively. In particular, the explicit
form of the metric and the warp factor have already been obtained in
\refs{\PWcrpt,\CLPopecrpt}. Let us first recall this result.

In the cartesian coordinates on $\IR^6$ with $S^5$ given as a unit
sphere, $\sum (x^I)^2=1$, the internal metric is:
\eqn\alchmetr{
ds_5^2(\al,\chi)\eql {a^2\over 2}{{\rm sech}\chi\over
\xi}(dx^IQ^{-1}_{IJ}
dx^J)+{a^2\over 2}{\sinh\chi\tanh\chi\over \xi^3} (x^IJ_{IJ}dx^J)^2\,.
}
Here $Q$ is a diagonal matrix
with $Q_{11}=\ldots=Q_{44}=e^{-2\alpha}$ and $Q_{55}=Q_{66}=e^{4\alpha}$,
$J$ is an antisymmetric matrix with $J_{14}=J_{23}=J_{65}=1$, and 
$\xi^2\eql x^IQ_{IJ}x^J$.   The  warp factor is simply
\eqn\warp{
\Omega^2\eql \xi~\cosh\chi\,. }
The constant $a$, introduced to account for an arbitrary normalization
of the Killing vectors, is fixed by requiring that at the ${\cal N}=8$
point the ten-dimensional metric becomes a product of $AdS_5\times
S^5$ with equal radii, $L=2/g$ and $a/\sqrt{2}$, respectively, which
gives $a\eql \sqrt{2} L$.

We need suitable spherical coordinates in which the $SU(2)\times
U(1)^2$ symmetry of the metric becomes manifest. First define complex
coordinates corresponding to $J$,
\eqn\complcoor{
u^1\eql x^1+ix^4\,,\qquad u^2\eql x^2+i x^3\,,\qquad u^3\eql x^5-ix^6\,,
}
and then reparametrize them using the group action\foot{Note that,
unlike in \PWcrpt, the $SU(2)$ doublet is inert under the $\phi$ rotation.}
\eqn\angcoords{
\left(\matrix{u^1\cr u^2\cr}\right)\eql
 \cos\theta\, g(\al_1,\al_2,\al_3)\,
\left(\matrix{1\cr 0\cr}\right)\,,\qquad
u^3\eql  e^{-i\phi}\,\sin\theta\,,}
where $g(\al_1,\al_2,\al_3)$ is an $SU(2)$ matrix expressed in terms of
Euler angles.

Define
\eqn\Xfac{\eqalign{
X_1(r,\theta)& \eql \cos^2\theta+\rho(r)^6\,\sin^2\theta\,,\cr
X_2(r,\theta)& \eql {\rm sech}\,\chi(r)\,\cos^2\theta+
\rho(r)^6\,\cosh\chi(r)\,\sin^2\theta\,,\cr}
}
By performing the change of variables \angcoords, we find that
\eqn\angxi{
\xi\eql {{X_1^{1/2}}\over \rho}\,,
}
and the ten-dimensional metric can be diagonalized in terms of the
following frames:
\eqn\theframessp{\eqalign{
e^{\mu+1} &\eql  {X_1^{1/4}(\cosh\chi)^{1/2}\over
\rho^{1/2}}\,e^A\,dx^\mu\,,\qquad
\mu=0,\ldots,3\,,\cr
e^5 &\eql {X_1^{1/4}(\cosh\chi)^{1/2}\over
\rho^{1/2}}\,dr\,,\cr
e^6 &\eql {2\over g} {X_1^{1/4}\over \rho^{3/2}\,
(\cosh\chi)^{1/2}}\,d\theta\,,\cr
e^7&\eql {1\over g}\,{\rho^{3/2} \,\cos\theta\over 
X_1^{1/4}\, (\cosh\chi)^{1/2}}\,\sigma_1\,,\cr
e^8&\eql {1\over g}\,{\rho^{3/2} \,\cos\theta\over 
X_1^{1/4}\, (\cosh\chi)^{1/2}}\,\sigma_2\,,\cr
e^9&\eql{1\over g}\,{\rho^{3/2}\, X_1^{1/4}\,\cos\theta\over
X_2^{1/2}}\,\sigma_3\,,\cr
e^{10}&\eql {2\over g}\,{X_2^{1/2}\,\sin\theta\over \rho^{3/2}
X_1^{3/4}}
\,d\phi+{1\over g}\,{\rho^{9/2}\,\sinh\chi\,\tanh\chi\,
\cos^2\theta\,\sin\theta\over X_1^{3/4}\,X_2^{1/2}}\,\sigma_3\,,\cr
}}
where $\sigma_i$, $i=1,2,3$, are the $SU(2)$ left-invariant 1-forms
normalized according to $d \sigma_i = {1\over 2} \epsilon_{ijk}\,
\sigma_j \wedge \sigma_k$.

An explicit evaluation of the dilaton/axion matrix \DilAnsatz\ yields
a somewhat surprising result: {\it the ten-dimensional dilaton/axion field
remains constant along this flow}.  This is surprising from the
field theory perspective in that one might have expected a running
coupling.  What we find is that the  dilaton and axion value provides
a modulus for the theory all along the flow: at the UV point this
is simply the gauge theory coupling constant, but in the infra-red
this presumably defines the line of marginal perturbations identified
in \RLMS.  

The constancy of the dilaton and axion is not so surprising from
the supergravity perspective:  The product structure of 
\scalmanLS\ means that the running fermion mass does
not mix with the five-dimensional dilaton/axion $SL(2,\IR)$
to produce a non-trivial ten-dimensional dilaton and axion.
This fact simplifies considerably the
ten-dimensional equations of motion.

Having exhausted the consistent truncation ansatz, our strategy is to
use the field equations and the underlying symmetry to construct the
remaining fields. As in \refs{\PWcrpt,\PWntwo}, we start with the
Einstein equations which should yield information about the field
strengths of the antisymmetric tensor fields given that the left hand
side, \ie\ the Ricci tensor, is computable.  The crucial observation
here is that the Ricci tensor depends on the derivatives of $A(r)$
only, and thus by using repeatedly the flow equations \floweqs\ and
\fieldeqs, one can eliminate all derivatives with respect to the flow
parameter, $r$, and be left with rational expressions in $\rho$ and
the hyperbolic functions of $\chi$. It is also reasonable to expect
that an explicit solution for trajectories (cf.\ \sersol) will involve
transcendental functions of $\cosh\chi$ and $\sinh\chi$ and thus we
should attempt to solve the ten-dimensional equations by matching
various $\rho$ and $\chi$ terms independently.  This is how
the lift worked for the ${\cal N}=2$ flow in \PWntwo, where the
explicit solution to the flow equations was not needed: 
the equations themselves were sufficient.

The resulting Ricci tensor is rather complicated to the extent that we
will not attempt reproducing it here. Nevertheless we find two simple
linear combinations that will become important in the following:
\eqn\rnodil{
R_{77}\eql R_{88}\eql R_{11}\,,}
and
\eqn\rofcomb{
R_{99}+R_{10\,10}-2 R_{11}\eql 2
g^2\,{\rho^3\,\sinh\chi\,\tanh\chi\over X_1^{1/2}}\,.
}
We also find that the only nontrivial off-diagonal components are
$R_{56}$ and $R_{9\,10}$.

The 5-index antisymmetric tensor field, $F_{(5)}$, is taken to be of
the similar form as for the ${\cal N}=2$ flow \PWntwo, namely
\eqn\theFten{
F_{(5)}\eql {\cal F}+*{\cal F}\,,}
where
\eqn\scrF{
{\cal F}\eql dx^0\wedge dx^1\wedge dx^2\wedge dx^3\wedge (w_r
dr+w_\theta  d\theta)\,,}
with  arbitrary functions $w_i(r,\theta)$. The self-duality
equation is then satisfied by construction. The structure of the 
energy-momentum tensor, $T^{(5)}_{MN}$, is the same as in 
\PWntwo, namely
\eqn\emFFten{
T_{11}^{(5)}\eql -T_{22}^{(5)}\eql\ldots\eql
-T_{33}^{(5)}\eql T_{77}^{(5)}\eql\ldots\eql
T_{10\,10}^{(5)} \eql {\cal A}^2+{\cal B}^2\,,}
\eqn\emFFtwo{
T_{55}^{(5)} \eql - T^{(5)}_{66}\eql {\cal A}^2-{\cal B}^2\,,}
and
\eqn\emFFthr{
T_{56}^{(5)}\eql T_{65}^{(5)}\eql 2 {\cal A}{\cal B}\,,}
where
\eqn\theab{\eqalign{
{\cal A}&\eql g\,e^{-4A}\, 
{\rho^{7/2} (\sech\chi)^{3/2}  \over X_1^{5/4}}\, w_\theta\, \,,\cr
{\cal B}&\eql - 2\, e^{-4A}\, {\rho^{5/2} (\sech\chi)^{5/2}\over X_1^{5/4}}
\, w_r \,.\cr} }

The most general $SU(2)\times U(1)$ symmetric Ansatz for the potential
$A_{(2)}$ of the antisymmetric tensor field is
\eqn\asymans{
A_{(2)}\eql e^{-i\phi}\, (a_1\,d\theta - a_2\sigma_3 - a_3\,d\phi)
\wedge (\sigma_1-i\sigma_2) \,,
}
where $a_i(r,\theta)$ are some arbitrary functions. This generalizes
the result in \PWcrpt, except for the $a_3$ term which, unlike in
\PWcrpt,  cannot be gauged away because of the $r$ dependence. 
Also the the $U(1)$ charge $-1$ is different than in \PWcrpt\ because
of the different $\phi$-dependence of the spherical coordinates
\angcoords.

In the absence of the dilaton/axion field, the 3-index antisymmetric
tensor field $G_{(3)}$ is simply $G_{(3)}=dA_{(2)}$. Since
$d(\sigma_1-i\sigma_2)=i(\sigma_1-i\sigma_2)\wedge \sigma_3$, we find
that $(\sigma_1-i\sigma_2)$ is a factor in $G_{(3)}$ so that
$G_{MNP}G^{MNP}=0$, as required by the dilaton equation. It also
implies that the  energy-momentum tensor, $T^{(3)}_{MN}$, satisfies
$T^{(3)}_{77}=T^{(3)}_{88}=T^{(3)}_{11}$. Then, given \emFFten, the
Einstein equations imply \rnodil, which provides us with the first
nontrivial test of the vanishing of the dilaton/axion.

Next we consider the solution to the linearized Einstein and Maxwell
equations at the UV end of the flow. From the diagonal Einstein
equations we recover, up to a sign, the usual Freund-Rubin Ansatz 
for the 5-index tensor,
\eqn\fivelim{
F_{12345}\eql F_{678910}\eql -{g\over 2}\,.
}
Substituting this into the Maxwell equations together with 
\eqn\linofG{
a_i(r,\theta)\eql e^{- r/L} \,\tilde a_i(\theta) 
+O(e^{-2 (r/L)})
\,, }
we look for a regular solution that also satisfies the (9,10) Einstein
equation. The latter does not involve the 5-index tensor and thus has
the lowest order contribution from the 3-index tensor. It turns out
that a required solution exists only for the choice of sign as in
\fivelim, and we we find
\eqn\linsola{
\tilde a_1(\theta)\eql {2\over g^2}\,\cos\theta\,,\qquad
\tilde a_2(\theta)\eql {1\over g^2} \,\cos^2\theta\,\sin\theta\,,\qquad
\tilde a_3(\theta)\eql -{2\over g^2} \,\cos^2\theta\,\sin\theta\,.
}

Turning to the general case we examine the combination 
\eqn\cruccomb{
T^{(3)}_{99}+T^{(3)}_{10\,10}-2\,T^{(3)}_{11}\eql
-{g^4\over 4}\,{\rho\,\cosh\chi\over X_1^{1/2}\, \cos^2\theta}\,
\left({\partial a_1\over \partial r}\right)^2+
{g^6\over 4}\, {X_1^{3/2}\,\cosh\chi\over \rho^3
 \,\cos^4\theta\,\sin^2\theta}\, (a_2-a_3)^2\,,
}
corresponding to \rofcomb.  The $\theta$-dependence on the right hand
side suggests that the functions $a_i(r,\theta)$ should be a simple
modification of their linearized conterparts, $\tilde a_i(\theta)$. In
particular,
\eqn\theaone{
{\partial a_1\over\partial r}\eql 
{1\over g}\, {(\rho^6-2)\,\tanh\chi\over \rho^2}\,\cos\theta\,,
}
and
\eqn\atwothr{
a_2-a_3\eql {1\over g^2}\,{(\rho^6+2)\tanh\chi\over X_1}\,
\cos^2\theta\,\sin\theta\,.
}
The first equation is simply integrated as
\eqn\ansaone{
a_1(r,\theta)\eql {2\over g^2}\,\tanh\chi\,\cos\theta\,.  }
Substituting \atwothr\ and \ansaone\ into the (9,10) Einstein equation
we finally determine that
\eqn\atwothre{\eqalign{
a_2(r,\theta)&\eql {1\over g^2}\, {\rho^6\,\tanh\chi\over X_1}\,
\cos^2\theta\,\sin\theta\,,\cr
a_3(r,\theta)&\eql -{2\over g^2}\, {\tanh\chi\over X_1}\,
\cos^2\theta\,\sin\theta\,,\cr}
}
and thus solve all the Einstein equations that do not involve the
5-index tensor.

The solution for the 5-index tensor is  easily  obtained from, 
for example, the (1,1), (5,5) and (5,6) Einstein equations, 
with any sign ambiguities
resolved by comparing the result with  the linearized limit. We find
\eqn\thetws{\eqalign{
w_r&\eql {g\over 8}\, e^{4 A}\, {\cosh^2\chi  \over \rho^4}\,
\big( (\cosh(2\chi)-3)\,\cos^2\theta + \rho^6\,(
2\rho^6\sinh^2\chi\,\sin^2\theta+\cos(2\theta)-3)\big)\,,\cr
w_\theta&\eql {
e^{4 A}\over 8\rho^2}\,\big(2\cosh^2\chi+\rho^6\,(\cosh(2\chi)-3)
 \big)\,\sin(2\theta)\,,\cr}}
and verify that
\eqn\intgrt{
{\partial w_r\over\partial\theta}\eql {\partial w_\theta\over\partial
r}\,, } which shows that $w_rdr+w_\theta d\theta=dw$ for some function
$w(r,\theta)$.

At this point all the fields have been determined and we verify that
all the remaning Einstein equations, the Maxwell equations and the
Bianchi identities are satisfied. 

\newsec{$\cN =2$ supergravity with hypermultiplets}

The truncations that we consider here are motivated by the flow
considered in \GPPZ.  The idea was to consider an $\cN=1$
supersymmetric flow in which all the chiral multiplets are given a
mass, leaving only the massless vector multiplet.  For simplicity, all
the masses are set equal and so the flow has an $SO(3)$ symmetry
rotating the three chiral multiplets into one another.  As we will
discuss below, the truncation of the supergravity to the $SO(3)$
invariant sector leaves eight scalar fields.  However, in \GPPZ\ only
two of these scalars were considered, and while these were the scalars
of physical interest, it was unclear as to whether they represented a
consistent truncation of the full set of eight.  It turns out that
it is indeed consistent to truncate to these scalars, and one way to
establish this is to show that they are the invariants under an
additional discrete symmetry.  We will also discuss this in some
detail below since this discrete symmetry has some interesting
consequences for the physics.

\subsec{The $SO(3)$ invariant sector}

The fermion mass matrix, and the corresponding supergravity
scalars can be represented as a complex, symmetric matrix
$m_{ij}$, $i,j = 1,\dots,4$.  The flow described in
\GPPZ\ involves setting $m_{ij} ={\rm diag}(m,m,m,0)$.  
The $SO(3)$ invariance is thus the orthogonal rotations on $a,b =
1,2,3$.  In particular it is the real subgroup of $SU(3) \subset SU(4)
= SO(6)$.  The ${\bf 4}$ of $SO(6)$ therefore decomposes as ${\bf
4\rightarrow 3 + 1}$ and ${\bf 6}$ decomposes as ${\bf 6\rightarrow
3+3 }$ of $SO(3)$. As mentioned earlier, the truncation to the
space of $SO(3)$  singlets reduces the $\cN =8$ supergravity theory to
$\cN=2$ supergravity coupled to two hypermultiplets, and the
scalar manifold is given by \scalmanQCD.

In terms of the Yang-Mills theory on the branes, the eight
scalars are dual the gauge coupling, the theta-angle, the
scalar operators:
\eqn\scalops{ \cO_1 \eql \sum_{j=1}^3~\big( {\rm Tr} \big( X^j 
X^j \big)  ~-~ {\rm Tr}\big( X^{j+3} X^{j+3}) \big) \,, \qquad  
\cO_2 \eql  \sum_{j=1}^3~\big( {\rm Tr} \big( X^j X^{j+3} \big)  \,, }
and the two complex fermion bilinears:
\eqn\fermibis{ \cO_3 \eql \sum_{a=1}^3~{\rm Tr} \big( \lambda^a 
 \lambda^a \big)   \,, \qquad  
\cO_4 \eql  {\rm Tr} \big( \lambda^4 \lambda^4 \big) \,. }
The coefficients of $\cO_3$ and $\cO_4$ are two complex, or
four real  parameters. 
One should also remember that the supergravity magically
adjusts the proper amount of 
$$\cO_0 \equiv \sum_{j=1}^6~ {\rm Tr}  \big( X^j  X^j\big)\,.$$

\subsec{Much ado about $G_{2(2)}$}

Our first task is to find an effective way to parametrize the manifold
\scalmanQCD.   Recall that
$E_{6(6)}$ has a maximal subgroup $SL(6,\IR) \times SL(2,\IR)_{5d}$.
Here we have put a subscript $5d$ on this $SL(2,\IR)$ to distinguish
it as that of the dilaton and axion of the ten-dimensional IIB theory.
The $SO(3)$ is the compact subgroup of the diagonal $SL(3,\IR) \subset
SL(3,\IR) \times SL(3,\IR) \subset SL(6,\IR)$.  The $G_{2(2)}$
subgroup of $E_{6(6)}$ that we seek in fact commutes with the diagonal
$SL(3,\IR)$.  Thus we have:
\eqn\SOembedd{
E_{6(6)}\supset SL(3,\IR)\times G_{2(2)}\supset SO(3)\times G_{2(2)}\,, }
for which
\eqn\decomp{{\bf 27\rightarrow (\overline 6,1)+(3,7)\rightarrow
(1,1)+(5,1)+(3,7)}\,. }
We now need to see how the invariances of the supergravity potential
act on this manifold.

First note that the diagonal $SL(3,\IR)$ commutes with 
$SL(2,\IR)_X$ in $SL(6,\IR)$, where the subscript $X$ is to 
distinguish  from $SL(2,\IR)_{5d}$.  
Hence, the $G_{2(2)}$ contains $SL(2,\IR)_X \times SL(2,\IR)_{5d}$. 
The non-compact generators of  $SL(2,\IR)_X$ are dual to
the operators $\cO_1$ and $\cO_2$ of \scalops.

Of the original $SO(6)$ invariance of the scalar potential, only 
the $SO(2)$ subgroup of  $SL(2,\IR)_X$ survives.  In addition, 
the potential is invariant under $SL(2,\IR)_{5d}$.  This four 
parameter family of invariances reduces the
eight-manifold to four independent parameters.  

There are, of course, many ways to parametrize the manifold, but the
simplest form that we have found is discovered by using the $SU(2)$ that
is diagonal\foot{Care is needed here since there is also an anti-diagonal
embedding, but this does not have the invariance structure that we need.}
in the denominator $SU(2)$'s of \scalmanQCD.  Under this $SU(2)$
the eight non-compact generators decompose as a ${\bf 5} + {\bf 3}$.
It turns out that the $O(2)$ subgroup of $SL(2,\IR)_X$  and the
non-compact generators of $SL(2,\IR)_{5d}$ can be used to set
the non-compact generators of the ${\bf 3}$ to zero, leaving the
${\bf 5}$.  Remarkably enough, this ${\bf 5}$ extends the $SU(2)= SO(3)$
to another $SL(3,\IR)$.  Thus we will parametrize the scalar potential
using this $SL(3,\IR)/SO(3)$.  There is still the residual invariance
of the compact generator of $SL(2,\IR)_{5d}$.  This acts on $SL(3,\IR)$
as a rotation in the first and second entries.
Explicit details of how this particular gauge choice is made in $G_{2(2)}$
may be found in Appendix A. (For another explicit parametrization of 
the coset, see \BodCad.)

Finally, to parametrize  $S \in SL(3,\IR)/SO(3)$, we will write it as
\eqn\Sparam{S \eql \cO(\theta_1,\theta_2,\theta_3)^{-1}~D~
\cO(\theta_1,\theta_2,\theta_3)  \,,\quad {\rm where} \quad
D~=~ {\rm diag}(\rho_1\,,\rho_2\,, ( \rho_1 \rho_2)^{-1}  )\,,}
and where $\cO(\theta_1,\theta_2,\theta_3)$ is a general $SO(3)$ rotation 
matrix.  It is usually convenient to parametrize such a rotation matrix using 
Euler angles, \ie\ by fundamental rotations, $R_{ij}(\theta)$, through
an angle $\theta$ in the $i$-$j$-plane:
\eqn\eulerangs{ \cO(\theta_1,\theta_2,\theta_3) \eql R_{12}(\theta_1)~
R_{23}(\theta_2)~ R_{12}(\theta_3) \,.}
If one uses this form of $\cO$ then the residual invariance
will mean that the potential is independent of the angle $\theta_3$.

\subsec{The scalar sector of the $\cN=2$ theory}

Using the parametrization described above, we find the 
following expression for the scalar potential of the 
$SO(3)$-invariant subsector:
\eqn\Gtwopot{\eqalign{\cP \eql & -{3 g^2 \over 8} ~-~
{3 \over 128}\,\big(\rho_1^2 - \rho_1^{-2} \big) \,
\big(\rho_2^2 - \rho_2^{-2} \big) ~-~ 
{3 \over 32}  \,\big( \rho_1^2  + \rho_1^{-2} + \rho_2^2 + 
\rho_2^{-2}\big) \cr & +~
{3 \over 128} \, \big(\rho_1 -  \rho_1^{-1} \big)^3\,
\big(\rho_1 \,\rho_2^2 - \rho_1^{-1}\, \rho_2^{-2} \big)  
\sin^2(\theta_2) \cr & -~ 
{3 \over 128} \big( \rho_1\, \rho_2^{-1} - \rho_1^{-1} \,
\rho_2 \big) \, \big(  \rho_1 \,\rho_2 -   \rho_1^{-1}\, 
\rho_2^{-1} \big)^3 \sin^2(\theta_1) \sin^2(\theta_2)  \,.}}

One can easily check that this potential yields no other
critical points other than the ones discovered in \KPW.

One of the key elements of five-dimensional supergravity 
is the matrix, $W_{ab}$, that appears in the
supersymmetry transformation of the gravitino \GRW. 
It is the eigenvalues of this matrix that generically provides
a superpotential in $\cN=1$ supersymmetric sub-sectors \FGPWa.
On the $SO(3)$ invariant $G_{2(2)}$ sector, we find that $W_{ab}$ 
consists of four two-by-two blocks, three of
which are identical. (This structure is required by $SO(3)$ invariance.)
The multiplicity-one block corresponds to the indices $(3,7)$ and
will be denoted ${\cal M}_1$, while the multiplicity-three block 
corresponds to the index pairs: $(1,5)$, $(2,6)$, $(4,8)$, and 
will be denoted ${\cal M}_2$.  Writing
\eqn\mmatr{
{\cal M}_j ~=~ \left(\matrix{ \alpha_j + i \beta_j  & -i \gamma_j \cr  
-i \gamma_j & \alpha_j - i \beta_j\cr} \right) \,, \quad j=1,2\,,
}
one has
\eqn\abcdefns{\eqalign{\alpha_1 ~=~ -{3 \over 8\, \rho_1^2 \rho_2^2}~
\Big[ & \rho_1 \rho_2 \big(1 + \rho_1^2 \big)\,
\big(1 + \rho_2^2 \big)\, ~+~ \rho_2 \big(1 -  \rho_1 \big)\, 
\big(1 + \rho_1^2 \big)\, \big(1 -   \rho_1 \rho_2^2 \big)\, 
\sin^2(\theta_2)   \cr  ~+~ & 
\big(\rho_1 - \rho_2  \big)\,\big(1 - \rho_1 \rho_2  \big)\,
\big(1 + \rho_1^2 \rho_2^2  \big)\,\sin^2(\theta_1) \sin^2(\theta_2)
\Big]  \,, \cr
\alpha_2 ~=~ - {1 \over 8\, \rho_1^2 \rho_2^2}~
\Big[ & \rho_1 \big(1 + 5 \rho_2^2 + 5 \rho_1^2 \rho_2^2 + 
\rho_1^2 \rho_2^4\big)\, \sin^2(\theta_1) \sin^2(\theta_2) \cr & ~+~ 
\rho_2 \big(1 + 5 \rho_1^2 +  5 \rho_1^2 \rho_2^2 + \rho_1^4 
\rho_2^2\big)\,\cos^2(\theta_1) \sin^2(\theta_2)
\cr & ~+~ \rho_1 \rho_2 \big(5 +  \rho_1^2 + \rho_2^2 + 
5 \rho_1^2 \rho_2^2 \big)\, \cos^2(\theta_2) \Big]  \,, \cr
\beta_1 ~=~  {3 \over 8\, \rho_1^2 \rho_2^2}~ &\big(\rho_1 - 
\rho_2  \big)\, \big(1 - \rho_1 \rho_2  \big)\, \big(1 + \rho_1^2 
\rho_2^2  \big)\, \sin (\theta_1) \cos(\theta_1)  \sin(\theta_2) \,, \cr
\beta_2 ~=~ - {1 \over 8\, \rho_1^2 \rho_2^2}~ &\big(\rho_1 - \rho_2  
\big)\, \big(1 - \rho_1 \rho_2  \big)\, \big(1 - 4 \rho_1 \rho_2 + 
\rho_1^2 \rho_2^2  \big)\, 
\sin (\theta_1) \cos(\theta_1)  \sin(\theta_2) \,, \cr
\gamma_1 ~=~  {3 \over 8\, \rho_1^2 \rho_2^2}~ \Big[ & 
\rho_2 \big(1 -  \rho_1 \big)\, \big(1 + \rho_1^2 \big)\, 
\big(1 -  \rho_1 \rho_2^2 \big)\, ~+~  \cr & \qquad 
\big(\rho_1 - \rho_2  \big)\, \big(1 - \rho_1 \rho_2  \big)\, 
\big(1 + \rho_1^2 \rho_2^2  \big)\,\sin^2(\theta_1)
\Big] \sin (\theta_2) \cos(\theta_2)   \,, \cr
\gamma_2 ~=~  {1 \over 8\, \rho_1^2 \rho_2^2}~ \Big[ & 
\rho_2 \big(1 -  \rho_1 \big)\, \big(1 - 4 \rho_1 + \rho_1^2 \big)\, 
\big(1 -  \rho_1 \rho_2^2 \big)\, ~+~  \cr & \qquad \big(\rho_1 - 
\rho_2  \big)\, \big(1 - \rho_1 \rho_2  \big)\, \big(1 - 
4 \rho_1 \rho_2 + \rho_1^2 \rho_2^2  \big)\,
\sin^2(\theta_1) \Big] \sin (\theta_2) \cos(\theta_2)   \,.
}}

The eigenvalues of $W_{ab}$ thus come in complex conjugate pairs with 
degeneracies $3$ and $1$, and are given by:
\eqn\evals{\lambda_j ~=~ \alpha_j~\pm ~i~\sqrt{\beta_j^2 + \gamma_j^2} \,,
\quad j= 1,2 \,.}
{}From previous experience, it is these eigenvalues that can give rise to
superpotentials, and in particular it is $\lambda_1$ that could potentially
be the superpotential for an $\cN=1$ theory.

The kinetic term of the subsector parametrized by $\rho_1, \rho_2$ and
$\theta_j, j=1,2,3$ is given by:
\eqn\Kterm{\eqalign{- &\coeff{1}{2}\, \big( (\del \varphi_1)^2 ~+~    
(\del \varphi_2)^2 ~+~  \, (\del \varphi_1) (\del \varphi_2) \big) ~-~ 
2\, \sinh^2(\varphi_1 -\varphi_2)\, ( \del 
\theta_1)^2 \cr & ~-~  4\, \sinh^2(\varphi_1 -\varphi_2)\, 
\cos\theta_2 \, ( \del  \theta_1)\,( \del  \theta_3)\cr & ~-~ 
2\,  \big(\sinh^2(\varphi_1 + 2 \varphi_2) + \sinh 
(\varphi_1 -\varphi_2) \sinh 3(\varphi_1 +\varphi_2)\, \sin^2\theta_1 \big) 
\,( \del  \theta_2)^2 \cr & ~+~ 
4\,  \sinh (\varphi_1 -\varphi_2) \sinh 3(\varphi_1 +\varphi_2)\, 
\sin\theta_1\, \cos\theta_1\, \sin\theta_2 
\,( \del  \theta_2) \, ( \del  \theta_3) \cr & ~-~ 
2\,  \big(\sinh^2(\varphi_1 - \varphi_2) +  \sinh 
(3 \varphi_1)\, \sinh(\varphi_1 +2\varphi_2)\, \sin^2\theta_2  
\cr & \qquad\qquad  -
\sinh (\varphi_1 -\varphi_2) \sinh 3(\varphi_1 + \varphi_2)\,  
\sin^2\theta_1\,  \sin^2\theta_2\big)  \,( \del  \theta_3)^2
\,,}}
where $\rho_1= e^{\varphi_1}$ and  $\rho_2= e^{\varphi_2}$.

The complexity, both literal and figurative, of the eigenvalues
\evals\ makes the isolation of $\cN=1$ superpotentials very difficult.
To facilitate this process, it is instructive to consider the field
theory duals of the supergravity scalars, and see how to reduce the
problem further.

\newsec{Further truncations of the $\cN=2$ theory}

Finding flows in the full set of scalars of the
$\cN=2$ theory is still rather difficult, and so
we simplify the problem further and reduce the
number of scalars by imposing discrete symmetries.

\subsec{The self-dual truncation}

The route taken in \GPPZ\ was to keep only $\cO_3$ and $\cO_4$ (and
implicitly $\cO_0$).  The corresponding supergravity scalars were
denoted by $m$ and $\sigma$ respectively, and the residual $U(1)$
invariances can be used to take $m$ and $\sigma$ to be real.  While
the results of \GPPZ\ are certainly correct, there were a few
omissions of detail, and as we will see at least one of these details
reveals some significant physics.

Setting $\theta_j =0$ and $\rho_1 = e^{{m\over\sqrt{3}}+\sigma } $, 
$\rho_2 =  e^{{m\over\sqrt{3}} - \sigma }$ in the paramtrization above
yields a diagonal $W_{ab}$ with one (multiplicity two) eigenvalue:
\eqn\GPPZWpot{W~=~ -\coeff{3}{4} \big(\cosh(2\, \sigma) + 
\cosh\big(\coeff{2 \, m}{\sqrt{3}}\big)\big)\,.}
The other eigenvalue is $   - {1\over 4} \big(\cosh(2\sigma)+
5 \cosh\big(\coeff{2m}{\sqrt{3}}\big)\big)$.
The potential \Gtwopot\ reduces to:
\eqn\potsimp{\cP \eql -\coeff{3 g^2}{16} \big[ 2 -  \coeff{1}{4}
\cosh (4 \sigma)  +   \coeff{1}{4}
\cosh \big(\coeff{4}{\sqrt{3}} m \big)  +   
\cosh \big(2 \sigma + \coeff{2}{\sqrt{3}} m \big)  +  
\cosh \big(2 \sigma  - \coeff{2}{\sqrt{3}} m \big) \big]\,,}
and the kinetic term takes the standard form:
\eqn\ktermsimp{ -  \coeff{1}{2}\,  (\del m)^2  -     
 \coeff{1}{2}\, (\del \sigma)^2 \,.}
One can easily check that:
$$
\cP \eql {g^2\over 8} \Big({\del W \over \del \sigma} \Big)^2  + 
{g^2\over 8} \Big({\del W \over \del m} \Big)^2 -  
{g^2\over 3} W^2 \,,
$$
and that there is no such equality for the other eigenvalue of
$W$.  

{}From this it is tempting to postulate \GPPZ\ that, as in \FGPWa,
$\cN=1$ supersymmetric flows are given by taking:
\eqn\floweqn{{d \varphi_j \over d r} ~=~
{g \over 2}~{\del W \over \del \varphi_j} \,,  \qquad  
 A' ~=~ - {g \over 3}~W \,,}
with $\varphi_1 =m$ and $ \varphi_2 = \sigma$.
However, to verify this one really needs to check the vanishing of
the supersymmetry variations of the spin-$\half$ fields: this we
have confirmed in detail.

One other detail that is not immediately apparent in \GPPZ\ is
the consistency of truncating to the $m$ and $ \sigma$ fields.  In
supergravity one might be concerned that the other fields of
the $G_{2(2)}$ coset do not decouple, while in field theory one
might be concerned that turning on $m$ and $ \sigma$ may cause other 
fields to flow.  Fortunately, explicit computation
reveals that \GPPZ\ is correct, and that this is a consistent
truncation, however it would be more satisfying if this fact were
understood as a result of a symmetry condition.  This is indeed possible.

Consider the following matrices:
\eqn\discsymm{\left( \matrix{0 & {\bf I}_{3\times 3} \cr 
-{\bf I}_{3\times 3} & 0}
\right) \,, \qquad \left( \matrix{0 & 1 \cr -1 & 0} \right) \,,}
where these are to be viewed as elements of $SL(6,\IR)$ and
$SL(2,\IR)_{5d}$ in  $SL(6,\IR) \times  SL(2,\IR)_{5d} \subset E_{6(6)}$.
These are invariances of the supergravity potential: indeed they
are elements of the invariance group $SO(6) \times SL(2,\IR)_{5d}$.
The simultaneous action  of these two matrices negates
the non-compact generators of $SL(2,\IR)_X$ and $SL(2,\IR)_{5d}$,
and leaves invariant precisely the (complex) parameters $m$
and $\sigma$.   Thus this  discrete symmetry  effects
the desired consistent truncation, and shows exactly why the 
other fields do not run in these models.

More significant  is the fact that this symmetry uses the modular inversion
of $SL(2,\ZZ) \subset SL(2,\IR)_{5d}$, combined with an $SO(6)$ rotation.
This should therefore be a symmetry of the underlying string theory as well,
and the invariance under \discsymm\ forces the UV string coupling, and
hence the Yang-Mills coupling on the brane to its self-dual value.
It is thus hard to see, from the field theory why the
super-QCD flow of \GPPZ\  should provide a model for electric and
magnetic confinement. 

To understand \GPPZ\ more completely, one should note that the modular
inversion is combined with a spatial inversion of $S^5$ in which the
first three cartesian coordinates, $x^1,x^2,x^3$, are exchanged with the 
second three, $x^4,x^5,x^6$.  This means that if one sees a characteristic
``electric  behavior'' by approaching on the $(1,2,3)$-axes then one must
be able to see the dual ``magnetic  behavior'' by approaching on the
$(4,5,6)$-axes.  As a result one sees that the confining behavior observed
in \GPPZ\ must be a pathology induced in Wilson and 't Hooft 
loops by approaching
the $S^5$ from a very special direction.  In reality an apparently
confining loop can lower its energy by slightly modifying its direction
of approach, and thereby become screened.   Thus the confining
behavior of \GPPZ\ is no more physical than that of \JMNW: it is
simply an artefact of an unstable symmetry axis.  This interpretation
is consistent with the analysis of \GPPZ\ in which the string tensions
were read off as eigenvalues of the $B$-field kinetic terms.  The
selection of an eigenvalue  is tantamount to selecting a direction
on $S^5$, and so the confining eigenvalues should be wiped out by
screening effects unless the $S^5$ is approached from very special 
directions \foot{ This last observation was made in
discussions with Joe Minahan.}.

\subsec{The parity-invariant sub-sector}

Before we leave the subject of consistent truncations, it is
worthwhile cataloging another potentially interesting subsector.
As we have seen, the flows of \GPPZ\ are self-dual, and it would be
nice to have a ``tame'' sector in which the five-dimensional
 dilaton could possibly flow.
One way to get such a sector is to require the parity symmetry:
\eqn\paritysymm{\left( \matrix{{\bf I}_{3\times 3} & 0\cr 
0 & -{\bf I}_{3\times 3}  }
\right) \,, \qquad \left( \matrix{-1 & 0 \cr 0 & 1} \right) \,.}
The former matrix lies in $O(6)$, and the latter is in  $GL(2,\IR)$. 
While the usual stated symmetry of the supergravity theory is
 $SO(6) \times SL(2,\IR)_{5d}$,  it is  actually symmetric under
$(O(6) \times SL^\pm(2,\IR)_{5d} )/\ZZ_2$, where $ SL^\pm(2,\IR)$ 
denotes the subset of $ GL(2,\IR)$ with determinant $\pm 1$,
and the division by $\ZZ_2$ requires that the determinants are equal
in each factor.  The parity symmetry \paritysymm\ projects out the 
operator $\cO_2$, the  five-dimensional 
axion, and enforces a reality condition on 
$m$ and $\sigma$.  This symmetry commutes with the $SO(3)$ symmetry, 
but it removes the
$U(1) \times U(1)$ symmetries of $SL(2,\IR)_X \times SL(2,\IR)_{5d}$.
We are thus left with four supergravity scalars, one of which is the 
five-dimensional dilaton.  

These four scalars turn out to be the non-compact directions of
yet another $SL(2,\IR) \times SL(2,\IR)$ in $G_{2(2)}$.  In terms
of the generators of Appendix A, these $SL(2,\IR)$'s are given by 
$L_1^{(1)} = -{1 \over 2} (X_4 + X_8),L_2^{(1)} = {1 \over 2} (X_1 + 
X_5), L_3^{(1)} ={1 \over 2}  (J_3 + K_3)$, and  
$L_1^{(2)} ={1 \over 2} (X_4 - X_8), L_2^{(2)} ={1 \over 2} (X_1 - X_5), 
L_3^{(2)} =-{1 \over 2}  (3 J_3 - K_3)$, with $L_1^{(j)}$  being compact.
We could proceed as above to  get at the scalar structure, but 
the previous parametrization did not handle the dilaton cleanly:
while it does not appear in the potential, the dilaton kinetic term
will mix in a complicated manner with the other scalars.  Here we
use a different gauge where the kinetic term is simple, but the
potential appears to depend upon all four scalars.  Each $SL(2,\IR)$
is parametrized using:
\eqn\SLtwocoords{\exp\big( -\phi_j L_1^{(j)}\big)~\exp\big( \alpha_j
 L_3^{(j)}\big)~ \exp\big(\phi_j L_1^{(j)}\big) \,, \quad j=1,2 \,.}
Define
\eqn\Wpotnew{\eqalign{W ~=~ & -{3\over 2}\,\cosh (\alpha_1 )\,\cosh 
(\alpha_2 )\, \big(  \cosh^2 (\alpha_2 )  - e^{4 i \,\phi_2}\, 
\sinh^2(\alpha_2 ) \big) 
\cr &~-~  {3\over 8} \,\sinh (\alpha_1 )\,\sinh (\alpha_2 ) \, 
\big( e^{2i\,(\phi_1  + \phi_2 )}+  3\, e^{2i\, ( \phi_1  + 
3 \phi_2)} \big) \cr &~+~ 
{3\over 4}\, i \,\sinh (\alpha_1 )\, \sinh (3\,\alpha_2 ) \,
\sin (2\,\phi_2 ) \, e^{2i\, ( \phi_1  + 2 \phi_2)}  \,,}}
then one has
\eqn\WPreln{ \cP ~=~ {1 \over 8}~\bigg| {\del W \over \del \alpha_1}
\bigg|^2 ~+~{1 \over 24}~\bigg| {\del W \over \del \alpha_2}
\bigg|^2 ~-~ {1 \over 3}~\big| W \big|^2  \,.}
Note that there are no derivatives of $W$ with respect to $\phi_j$
on the right-hand side of this equation. There is also an 
asymmetry between $\alpha_1$ and $\alpha_2$ in \Wpotnew\ because
the two $SL(2,\IR)$'s are different: the non-compact generators
of $G_{2(2)}$ form a $\bf (2,4)$ of these two groups.

In this parametrization the kinetic term takes the form:
\eqn\simpKterm{ - \coeff{1}{2}\, (\del \alpha_1)^2 -    
 \coeff{3}{2}\, (\del \alpha_2)^2 -   
2 \sinh^2(2\, \alpha_1) \, (\del \phi_1)^2 -    
6 \sinh^2(2\, \alpha_2) \, (\del \phi_2)^2 \,.}

\newsec{The metric and dilaton background}

\subsec{ The metric}

As desribed in the introduction, the 
ten-dimensional background metric is given
by the warped product \warpmetr, with an
internal metric on the deformed $S^5$ given by
\metanswer.    It is natural to use the 
$SO(3)$ in describing the internal geometry, and indeed
we will describe the deformed $S^5$ in terms of
a degenerate fibration of $SO(3)$ over the space
of its orbits.    We will also only consider the
metric corresponding to the two scalar subspace of
section 5.1 and   \GPPZ.

We start by thinking of the 
``round''  $S^5$ as the unit sphere in $\IR^6$, but with
 the cartesian coordinates split into two groups of 
three: $(u^i, v^j)$, $i,j= 1,2,3$. The $SO(3)$ symmetry acts 
upon these simultaneously in the vector representation.  The 
internal $5$-manifold  is still the unit sphere:
\eqn\uvsurf{ u^2+v^2 ~=~ 1\,, }
but the general metric on this deformed $S^5$ is given by:
$$
ds_5^2 \eql \xi^{-{3\over 2}}~ d \hat s_5^2  \,,
$$
where
\eqn\intmetr{\eqalign{ 
d \hat s_5^2 \eql&
a_1\,du^idu^i + 2\,a_2 \,du^i dv^i+ a_3 \, dv^i d v^i \cr
& +a_4 \, \big(d(u\cdot v)\big)^2 +  2\, a_5 \,(v^i du^i)(v^j du^j)
+ 2 \, a_6\, (u^i du^i)(v^j dv^j)\,.
\cr} }
The coefficient functions are then given by:
\eqn\acoefs{\eqalign{   a_1 ~=~ & {1 \over 4\, \mu^2\, \nu^4} ~
(1+  \mu^2\, \nu^2)~\big((1+  \mu^2\, \nu^2)\,\nu^2 \, u^2 ~+~
 (\mu^2 + \nu^6) \,  v^2 \big) \,, \cr
  a_2 ~=~ &- {1 \over 4\, \mu^2\, \nu^4} ~
(1 - \nu^4)~(1 - \mu^2\, \nu^2)~( \mu^2 + \nu^2)~ u \cdot v  \,, \cr
  a_3 ~=~ & {1 \over 4\, \mu^2\, \nu^4} ~
(1+  \mu^2\, \nu^2)~\big((\mu^2 + \nu^6) \, u^2 ~+~
 (1+  \mu^2\, \nu^2)\,\nu^2  \, v^2   \big) \,, \cr
  a_4 ~=~ & {1 \over 16\, \mu^4\, \nu^6} ~
(1 - \mu^2\, \nu^2)^2 ~ (1+ \mu^2 \nu^2)~(\mu^2 + \nu^6)  \,, \cr
  a_5 ~=~ & {1 \over 8\, \mu^4\, \nu^4} ~
(1 - \mu^4\, \nu^4) ~ (\mu^4 - \nu^4)  \,, \cr
  a_6 ~=~ &- {1 \over 8\, \mu^2\, \nu^6} ~
(1 -  \nu^8) ~ (\mu^4 - \nu^4)   \,,\cr
}}
where
\eqn\munudefns{ \mu \equiv   e^{\sigma }    \ , \qquad 
\nu \equiv e^{{m\over\sqrt{3}} }  \ .}

The warp-factor, $\xi$, is given by:
\eqn\warpfactor{\eqalign{\xi^2 ~=~ {1 \over 16\, \mu^4\, \nu^8}~
\Big[ & \nu^2\, (1 + \mu^2\, \nu^2)^3 ~(\mu^2 + \nu^6) ~+~ 
 (1 -  \nu^4)^2~(\mu^2 - \nu^2)^2 ~(1 + \mu^2\, \nu^2)^2~
u^2 \, v^2 \cr
  & -(1 - \mu^2\, \nu^2)^2~(1 -  \nu^4)^2~(\mu^2 + \nu^2)^2 ~
(u \cdot v)^2 \Big]\,.}}

Note that at $\mu=\nu =1$ the internal metric given by \intmetr\ 
and \acoefs, on the surface \uvsurf,  collapses to that
of the round sphere of unit radius.  Moreover, at $\mu=\nu =1$  one has
$\xi = 1$.  As usual, we define $\Delta$ by $\Delta^2 \equiv {\rm
det}\,( g_{mp}\,{\displaystyle \gop}{}^{pq})$, where $g_{mp}$ is the
internal metric on $S^5$ given by \intmetr\ and 
${\displaystyle \gop}{}^{pq}$
is the inverse of the ``round'' internal metric at $\mu=\nu=1$.
We then have 
\eqn\xiDelta{\xi ~\equiv~ \Delta^{-{4 \over 3}} \,,}
and the complete ten-dimensional metric is: 
\eqn\newwarpmetr{
ds_{10}^2\eql \xi^{1 \over 2} ~ds_{1,4}^2 ~+~ 
\xi^{-{3\over 2}}~ d \hat s_5^2\,.}

The foregoing metric on $S^5$ is far from elementary, but a 
natural way to think of it is as an $\IR \IP^3$ fibered over   
$\IP^1/(\ZZ_2 \times \ZZ_2)$.  The $\IR \IP^3$ fiber is,
of course,  $SO(3) \equiv S^3/\ZZ_2$, and the base is the 
the orbit space of this $SO(3)$ on $S^5$.  This base has the
topology of a disk.

To see this explicitly one can use the $SO(3)$ action to 
reduce $u$ and $v$ to:
\eqn\redgauge{u ~=~\left(\matrix{0 \cr 0 \cr u_3} \right) ~=~
\left(\matrix{0 \cr 0 \cr \cos \theta} \right) 
\,, \qquad v ~=~\left(\matrix{0 \cr v_2 \cr v_3} \right) ~=~
\left(\matrix{0 \cr \sin\theta~\sin \phi \cr
\sin\theta~\cos \phi} \right)   \,. }
The remaining non-zero elements satisfy $(u_3)^2 + (v_2)^2 + 
(v_3)^2 =1$, and so naively describe an $S^2$.  However,
any two coordinates can be negated by an $SO(3)$ rotation
and so we divide by the inversions: $v_2 \to - v_2$ and
$u_3 \to - u_3, v_3 \to - v_3$.  Thus we can make the
restrictions $v_2 \ge 0$, $ u_3 \ge 0$.  In terms of the 
polar coordinates, one has:  $0 \le \theta \le \pi/2$,
$ 0 \le \phi \le \pi$.  It should also be noted
that for $\theta = \pi/2$ the coordinate $\phi$
becomes redundant.  Given the  $\ZZ_2$ identifications
for general $\theta$ and $\phi$, the base  may be thought of as
a quarter sphere, which has the topology of a disk.  
We can parametrize it in terms of the coordinates:
\eqn\newcoords{w_1 = 2 \, u\cdot u - 1 = \cos(2 \theta) \,, 
\quad w_2 =  2 \, u\cdot v =   \sin (2\theta)~ \cos\phi \,,
\quad 0 \le w_1^2 + w_2^2 \le 1\,.}

The fiber is regular except at the edges of the disk,
\ie\ at points where
\eqn\singpts{w_1^2 ~+~ w_2^2 \eql 1  \qquad \Leftrightarrow 
\qquad \sin\theta\,\cos\theta\, \sin\phi = 0 \,.}
At these points  either $u$ or $v$ vanishes, or $u$ and 
$v$ are colinear.    At such points the
fiber degenerates to a $\IP^1$.  
We should stress that even though this description as an
$\IR \IP^3$ fibration is  singular, the overall manifold
at generic values of $\mu$ and $\nu$ is still a perfectly
smooth, but deformed, $S^5$.

\subsec{ The dilaton}
 
Using the scalar fields of section 5.1, we computed the
right-hand side of \DilAnsatz\ with $x^I= (u^i,v^j)$.  
There is an important
consistency check in that taking the determinant on both sides
of \DilAnsatz\ must give the same expression for $\Delta$ 
as that given by \xiDelta\ and \warpfactor. This does indeed work.  
Furthermore, we obtain the following components
for $\cM = S\,S^T$:
\eqn\Mdil{\eqalign{ \cM_{11} \eql & { 1 \over 4\, \xi \, 
\mu^2\, \nu^4}~ 
(1 + \mu^2\, \nu^2) \,\big( (\mu^2 + \nu^6)\, \cos^2\theta + 
\nu^2 (1+ \mu^2 \nu^2)\, \sin^2\theta \big) \,, \cr
\cM_{12} \eql & \cM_{21} \eql { 1 \over 4\,\xi\, \mu^2\, \nu^4}~
(1 - \nu^4)\,  (1 - \mu^2\, \nu^2) \, (\mu^2 + \nu^2)\sin\theta \, 
\cos \theta\, \cos\phi \cr
\cM_{22} \eql & { 1 \over 4\,\xi\, \mu^2\, \nu^4} ~ 
(1 + \mu^2\, \nu^2) \,\big(\nu^2 (1+ \mu^2 \nu^2)\,\cos^2\theta + 
 (\mu^2 + \nu^6)\, \sin^2\theta \big) \,. \cr   
 }}

Thus we have an extremely non-trivial dilaton/axion background.
Intriguingly enough, the matrix elements of $\cM$ are, up to a 
factor of $\xi$, exactly the same as the metric coefficients 
$a_1, a_2$ and $a_3$.  Thus the dilaton is controlling the relative
sizes of the $u$-sphere and $v$-sphere, while the axion controls
the fibering of one over the other.

\newsec{The flows and their infra-red asymptotics}

The mathematics of the flows were thoroughly described in 
\GPPZ, and we first summarize these results in our conventions.
The solution to \floweqn\ for $\mu = e^{\sigma}$ and $\nu = 
e^{{2 \over \sqrt{3}} m}$ is:
\eqn\flowsoln{ 
\mu ~=~  \sqrt{1 + \lambda t^3 \over  1- \lambda  t^3} \,, 
\qquad \nu ~=~ \sqrt{1 +t \over  1-t} \,, }
and 
\eqn\flowsolnA{
A(r) \eql  
\coeff{1}{6} \log\big( t^{-3} -\lambda^2 \, t^3 \big) ~+~
\coeff{1}{2} \log\big(t^{-1} -t\big) ~+~ C_1\,,}
where
\eqn\paramsol{t \eql \exp\big[ -\big({r\over L} - C_1\big) \big] \,, 
\qquad \lambda \eql \exp \big[3\big(C_2 -C_1\big)\big]  \,,}
and where the $C_j$ are constants of integration for the 
flows of $m$ and $\sigma$.  Indeed, near the UV limit one has:
\eqn\UVasymp{\eqalign{m ~\sim~ & m_0 e^{-{r \over L}} \,, 
\quad \sigma ~\sim~  \sigma_0\, e^{-3{ r \over L}} \,,\cr
m_0 ~\equiv~ &
\coeff{\sqrt{3}}{2}\, e^{C_1} \,, \quad  \sigma_0 ~\equiv~  
\coeff{1}{3}\, e^{3 C_2}\,, \quad \lambda \eql
\coeff{9 \sqrt{3}}{8}\, {\sigma_0 \over m_0^3} \, .}}
Thus the constants of integration therefore represent the values 
of the  mass and gaugino consensate introduced in the UV theory.
The constant of integration in
$A(r)$ has been chosen so that $A(r) \sim{r\over L} + O(e^{-r/L})$
as $r \to \infty$.  It was argued
in \GPPZ\ that the physical flows have $\lambda \le 1$, 
and thus have the fermion mass scale greater
that the gaugino condensate scale.

\subsec{Asymptotics for $\lambda < 1$}

For $\lambda < 1$ the five-dimensional metric becomes 
singular at $r = C_1 L$, or at $t=1$ \GPPZ.  The ten-dimensional
metric is, however, much less singular, and indeed resolves
into a ring distribution of what appear to be $7$-branes.  
To see this we  start by parametrizing the vectors $u,v$ by:
\eqn\uvparam{u ~=~\cR~\left(\matrix{0 \cr 0 \cr \cos \theta} \right) 
\,, \qquad v ~=~\cR~\left(\matrix{0 \cr \sin\theta~\sin \phi \cr
\sin\theta~\cos \phi} \right)  \,, \qquad 0 \le \theta \le \pi/2
\,, \  0 \le \phi \le \pi \,,}
where $\cR$ is a generic $SO(3)$ rotation matrix.  
One then decomposes $R^{-1} dR$ into the left invariant
$1$-forms, $\sigma_i$, $i=1,2,3$, normalized according
to $d \sigma_i = {1\over 2} \epsilon_{ijk}\, \sigma_j \wedge 
\sigma_k$.

In the limit $t \to 1$ the warp factor, $\xi$, diverges 
according to:
\eqn\xidiv{ \xi ~\sim~ {1 \over (1-t)^2}~(\sin \theta\,
\cos\theta\, \cos\phi)^{1\over 2} \,.}
The factor  of $\xi^{1/2}$ in \newwarpmetr\ makes two important
modifications to the five-dimensional metric.  First, it
exactly cancels the vanishing of $e^{2 A}$ as $t \to 1$, and
leaves a finite coefficient.  Secondly, it suggests the
change of variable $\chi \equiv 2 (1-t)^{1/2}$ to regularize
the radial behaviour.  

The net result of this is that the ten-dimensional metric
has the following leading behaviour in $\chi$ as $\chi \to 0$.
\eqn\asympmet{ \eqalign{ds^2 = & \coeff{L^2}{\sqrt{2}}
(1- w_1^2 -w_2^2)^{1\over 4}~ \bigg[ \coeff{2} {L^2} \,
( 1- \lambda^2)^{1/3} \,
e^{2C_1} \,\big(\eta_{\mu \nu} dx^\mu dx^\nu \big) - d \chi^2 -
\coeff{1}{4}\,\chi^2\, (\sigma_1^2 + \sigma_2^2  +\sigma_3^2 ) \bigg] 
\cr & ~-~ \coeff{L^2}{\sqrt{2}} \, (1- w_1^2 -w_2^2)^{-{3\over 4}}~ 
\bigg[ {2(1- \lambda) \over  (1+ \lambda)}~dw_1^2   ~+~ 
{(1+ \lambda) \over  2(1- \lambda)}~dw_2^2   \bigg]\,.}}
Observe that the metric in the first square bracket is
{\it locally} that of a flat Lorentzian $7$-brane, while the
metric in the second square bracket is that of a flat 
Euclidean metric on the disk.  The warp factor:
\eqn\warpfactortwo{\zeta ~\equiv~ 1 -w_1^2 - w_2^2  ~=~ 
4 \sin^2 \theta\, \cos^2 \theta \, \sin^2 \phi \,,}
is only singular on the ring at the edge of the disk.
Moreover the powers of $\zeta$ that appear in \asympmet\
are precisely those appropriate to a dimensional
reduction of ten dimensional physics to 
$(7+1)$-dimensional physics on the brane.  Thus we see
that in the IR limit the D3-brane physics 
appears to be oxidizing to $7$-brane physics.  

To be more explicit, far from the brane one sees the usual $D3$-brane
throat, but as one approaches $t=1$, or $r= C_1 L$, the throat rounds
out into a seven-brane world. Now recall that $e^{2 C_1} = {4 \over 3}
m_0^2 $ (see \UVasymp), where $m_0$ is the mass of the chiral
multiplets.  Thus the distance that one descends down the throat
before encountering the $7$-brane is set by the UV mass, $m_0$.  Also
note that the scale in front of the $D3$-brane metric is $( 1-
\lambda^2)^{1/3} \, e^{2C_1}$, and so the supergravity description of
this flow terminates at a $D3$-brane scale determined by the chiral
multiplet mass and by the gaugino condensate.  The larger the chiral
multiplet mass, the closer to the UV it terminates, but the nearer
$\lambda$ is to $1$, the nearer the IR the flow goes.

The seven-brane form of the metric is precisely consistent with 
the infra-red limit of the dilaton.  As $t \to 1$ the matrix $\cM$ in
\Mdil\ limits to:
\eqn\Mdillim{\cM~=~ {1 \over \sin \phi}~ \left( 
\matrix{ \cot\theta & \cos\phi \cr \cos\phi &
\tan\theta} \right)  \,.}
This dilaton/axion configuration is regular everywhere
except exactly where \warpfactortwo\ vanishes.

There is also an interesting  topological issue:  
while the first metric factor
in  \asympmet\ is locally flat, it is actually
$\IR^{3,1}\times \IR^4/\ZZ_2$ where the $\ZZ_2$ negates four of the 
spatial coordinates.  It thus has an $A_1$ singularity.
The reason for this is that the apparently spherical
section of the metric \asympmet\ represented by the left invariant 
one-forms, $\sigma_j$, is the metric on $SO(3) = S^3/\ZZ_2$ and
not the metric on  $SU(2) = S^3 $.  This is the origin of
the modding by $\ZZ_2$.  

This suggests that the
string theory will see new massless states associated
with branes wrapping this vanishing $2$-cycle.

\subsec{Asymptotics for $\lambda = 1$}

If one looks at \asympmet\ one sees that
various coefficients either vanish or diverge as 
$\lambda \to 1$.  In a more careful treatment of
the asymptotics these coefficients are, respectively, replaced
by positive or negative powers of the radial coordinate $\chi$.
To be more explicit, first note that the five-dimensional metric 
\RGFmetric\ now behaves according to:
\eqn\singmet{ ds^2_{1,4} = (1-t)^{4/3}~e^{2 C_1}~2^{4 \over3}\,
3^{1 \over 3} \,\eta_{\mu\nu} 
dx^\mu dx^\nu -  dr^2 \,.}
The warp factor is now asymptotic to:
\eqn\singwarp{\xi ~\sim~  {1 \over (1-t)}~ \widehat \Omega \,,
\quad {\rm where} \quad \widehat \Omega \equiv  
\coeff{1}{3}\,\big(3\, \cos^2(2 \theta) + 4\, \sin^2(2 \theta)\,
\sin^2(\phi) \big)^{1/2} \,.}
Once again one introduces the change of variables: 
$\chi \equiv 2 (1-t)^{1/2}$, and one then finds that the
ten-dimensional metric takes the form:
\eqn\singasympmet{ds^2 ~\sim~ \widehat \Omega^{1\over 2}~ 
\Big[  2^{2 \over 3}\, 3^{1 \over 3} \, e^{2C_1}\, 
\chi^{2 \over 3} \, \big(\eta_{\mu \nu} dx^\mu dx^\nu 
\big) - L^2\, d \chi^2 \Big] ~-~ L^2\,\widehat 
\Omega^{-{3\over 2}}~\Big[ ~{1 \over 3 \chi^2 }\, dw_2^2 ~+~  
 d\tilde s_4^2  \Big]  \,,} 
where $ d\tilde s_4^2$ is a complicated, but regular metric
on $\IR\IP^3$ and in the $\theta$ direction.

The dilaton matrix, ${\cal M}$, takes the form
\eqn\redgauge{{\cal M} ~\sim~
\coeff{1}{3}\,\widehat \Omega^{-1}~ \left(
\matrix{1 + 2 \cos^2(\theta) & 2\, \sin(2 \theta)\, \cos\phi \cr 
 2\, \sin(2 \theta)\, \cos\phi & 1 + 2 \sin^2(\theta)} \right)  \,. }

The metric and the dilaton no longer have a ring singularity,
but only have a singularity at the points 
$\theta = \pm {\pi \over 4}, \phi = 0$.  On the other hand, the
metric now has a singularity at $\chi =0$.  It is not so
simple to give this metric a geometric interpretation,  
particularly since one
of the internal directions is blowing up as $\chi \to 0$. 
On the other hand, in 
contradistinction to the $\lambda < 1$ flows, the 
$D3$-brane coefficient vanishes as $\chi \to 0$, which, in
principle, suggests that the flow might be able to probe further 
into the  infra-red.

Interestingly enough, the metric and dilaton becomes a little
more regular near the apparently singular region 
$\theta = \pm \pi/4, \phi =0$.  Setting 
$\theta =  {\pi \over 4} + \psi$, $t = 1 - {1 \over 2} \chi^2$ 
and expanding in small $\chi, \psi$ and $\phi$ we find:
\eqn\vsingasympmet{\eqalign{ds^2 ~\sim~ & 2\, \widetilde 
\Omega^{1\over 2}~  \Big[  3^{1 \over 3} \, e^{2C_1}\, 
\chi^{2 \over 3} \, \big(\eta_{\mu \nu} dx^\mu dx^\nu 
\big) - L^2\, d \chi^2 \Big] \cr &  ~-~ 
\coeff{1}{2\, \sqrt{3}} \, L^2 \, \widetilde 
\Omega^{-{3\over 2}}~\chi^2~\Big[ ~ \coeff{16}{3}\, d\psi^2 ~+~  
d\phi^2 + \phi^2\, (\sigma_1^2 +  \sigma_2^2 +\sigma_3^2 ) ~
\Big]  \,,}} 
where
\eqn\vsingwarp{  \widetilde \Omega \equiv  
\coeff{2}{3}\,\big(3\, \psi^2  + \phi^2 + \chi^4 \big)^{1/2} \,.}
Note that the metric \vsingasympmet\ has  round 
$\IR \IP^3$ fibers, but there is a conical singularity at $\phi =0$
\foot{In our conventions the   non-conical metric would be:
$d\phi^2 + {1 \over 4}\phi^2\, (\sigma_1^2 +  \sigma_2^2 +
\sigma_3^2 )$}.  The dilaton matrix becomes:
\eqn\vredgauge{{\cal M} ~\sim~\coeff{2}{3} \,\widetilde 
\Omega^{-1} ~\cQ~ \left( \matrix{2 & -\psi \cr  -\psi &  
2 \psi^2 +\coeff{1}{2} \,\phi^2 + \coeff{1}{2} \,\chi^4} \right)~\cQ^T 
\,, }
where $\cQ$ is a rotation by $\theta = \pi/4$.

\newsec{The ring singularity:  Looking for $5$-branes}

One of the motivations of \PolStr\ was to relate the
supergravity flows to
the non-commutative geometry suggested by the Yang-Mills
superpotential.  In particular, it was shown in \refs{\CVEW,
\RDEW} that the chiral superfields of the supersymmetric vacuum 
must obey:
\eqn\VWvac{\big[\Phi_i\,,\Phi_j\big] \eql -{m \over 
\sqrt{2}} \epsilon_{ijk}~ \Phi_k \,.}
Since these are the commutation relations of
$SU(2)$, the possible vacua are classified by the maps 
of $SU(2)$ into the gauge group $SU(N)$.   Indeed, if the
mass parameter, $m$, is real then the only solution to
\VWvac\ is to take $\Phi_j$ to be some real combination of the
anti-hermitian generators of $SU(N)$.  It was thus argued in
\refs{\Myers,\PolStr} that to find a ground state of the $\cN=1$ theory,
only  the real part of $\Phi_i$ can develop a vev, and then given that
$\sum_j Tr (|\Phi_j|^2 )\sim m^2$, it follows that the
vacuum state of the $\cN=1$ theory should correspond
to the $D3$-branes becoming dielectric $5$-branes that 
wrap  a non-commutative $S^2$.

To connect this with the results here, recall that for
finite $N$ and for commuting vevs, the $\Phi_j$ may be 
thought of as the cartesian coordinates transverse to the
$D3$-branes.  More generally, the solutions here have
an $SO(3)$-invariance:  in \VWvac\ the (real) $SO(3)$ acts 
the indices $i,j,k$, with the real and imaginary parts of $\Phi_i$
transforming separately, each as a triplet of $SO(3)$.  Thus the
real and imaginary parts of $\Phi_i$ correspond to the 
coordinates $u_i$ and $v_i$ on $S^5$.  If we were to obtain
precisely the solution of \PolStr\ then
the  $5$-branes should emerge in the limit in which 
$v_j \equiv 0$:  Instead we find a ring singularity  
when $\vec u$ and $\vec v$ are parallel. 

The key to understanding this apparent discrepancy
comes from looking at the flows  with $\lambda =0$.  In 
supergravity these flows have an additional $U(1)$ symmetry 
that is generated by the simultaneous action of the 
matrices \discsymm\ considered as $SO(2)$ generators
in $SL(6,\IR) \times SL(2,\IR)$.  This symmetry rotates
$\vec u$ into $\vec v$ while performing an ``S-duality
rotation'' in the $SL(2,\IR)$.  Because  this symmetry
is embedded partially in the $SL(2,\IR)$, this $U(1)$ will
not be a symmetry of the field theory at finite $N$:
at best it will reduce to some discrete subgroup
of the $SL(2,\ZZ)$, S-duality symmetry of the finite
$N$, $\cN=4$ Yang-Mills theory.  

Returning to
our ring singularity, one sees that it is essentially
given by $\phi =0$, and  $-\pi/2 \le \theta \le \pi/2$.   
Note that we have doubled the range of $\theta$ used in
\uvparam:  This is enables us to set $\phi=0$ and still cover 
the region with $u$ and $v$ anti-parallel ($\phi = \pi$).
This range of $\theta$ thus covers the whole ring
singularity.  On this locus, the $SO(2)$ action is 
represented by a rotation in $\theta$, and hence 
the symmetry sweeps out the ring. It follows that as we 
go around the ring, the singularity must be undergoing a
continuous ``S-duality rotation'' in $SL(2,\IR)$.  

This picture is confirmed by a more careful analysis of
the behaviour of the dilaton near the singularity.
In the previous section we took the limit in which  
$(1- w_1^2 -w_2^2)$ was finite, and 
$(1-t)$ was becoming vanishingly small, that is, we 
considered a generic {\it interior} point of the disk defined 
by $w_1$  and $w_2$.  The asymptotic behaviour of the metric
and dilaton depend upon the order of these limits, and we
now consider them in the opposite order.  We
will also restore $\lambda$, but keep $\lambda <1$.
The dilaton matrix now has the asymptotic form:
\eqn\Mring{\cM \eql Q \cdot \left(\matrix{\cU\, {1 \over 
\sqrt{1-t}}  & 0 \cr 0 & \cU^{-1} \, \sqrt{1-t}}
\right) \cdot Q^{-1} \,,}
where
\eqn\QUdefns{ \cU ~\equiv~ \Big( {2(1-\lambda^2)  
\over 1+2 \,\lambda \, 
 \cos(4\,\theta) + \lambda^2 } \Big)^{1 \over 2} 
\,, \qquad
Q ~\equiv~ \left(\matrix{\cos \theta & -\sin \theta \cr \sin \theta 
&\cos \theta } \right) \,.}  
Note that for $\theta =0$ ($v=0$)
and $\theta =\pm \pi/2$ ($u=0$) this dilaton configuration
is precisely that which is appropriate for  $NS$ $5$-branes
and $D5$-branes, and that these limits are exchanged
by the $\pi/2$ rotation corresponding to $\tau \to -1/\tau$.
In between we have a continuous rotation
$Q \in SO(2) \subset SL(2,\IR)$, and this is precisely the 
same as the rotation between   $\vec u$ and $\vec v$
that takes us around the ring.   

It is also instructive to parametrize the $SL(2,\IR)$ 
matrix in terms of the coupling, $\tau$.  One then finds:
\eqn\taueqn{\tau ~=~ {i \,\cU\, \cos \theta ~-~ \sqrt{1-t}\, 
\sin \theta \over \sqrt{1-t} \cos \theta ~+~i\,\cU \,  \, 
\sin \theta} ~\sim~ \cot \theta \quad {\rm as} \ t \to 1\,.}
As one goes around the ring one finds that the coupling
runs from infinity down to zero along the positive real
axis.  At finite $N$, a singularity
at $Im(\tau) = 0$ can be interpreted in terms 
$(p,q)$-branes provided that $\tau$ approaches
a rational point on the real axis.  It is only in the
limit $N \to \infty$ that we can get a smooth
distribution $(p,q)$-branes.

One can also  analyse the metric in the limit in which 
$(1- w_1^2 -w_2^2)$ vanishes faster than $(1-t)$, and one
sees qualitatively different behaviour from \asympmet.
There is also a hint
of the $5$-branes wrapping an $S^2$ \refs{\Myers,\PolStr}.  Indeed, 
it should  be recalled that in our description of the $S^5$ geometry,
the $\IR \IP^3$ fiber  degenerates to an $S^2$ on the ring 
singularity, and this is the $S^2$ upon which the $5$-branes
must wrap.  Again we focus upon the flows with $\lambda <1$.
We consider the metric near $t=1$ but with $\phi =0$ in
\uvparam.  The residual directions are thus the radial,
or $t$ direction, the $D3$-branes, the $S^2$ fiber, and
the angle, $\theta$.  We find:
\eqn\dseight{\eqalign{ ds_8^2 ~\sim~  \Big( {1+2 \,\lambda \, 
\cos(4\,\theta) + \lambda^2
\over 2(1-\lambda^2)} \Big)^{1 \over 4}\,  \Big( & -
(1-t)^{-{3 \over 4}}  \,  (dr^2 + d \theta^2) 
~-~ (1-t)^{+{5 \over 4}}  \, d\Omega_2^2 \cr & ~+~2 \, 
( 1- \lambda^2)^{1/3} \, (1-t)^{+{1 \over 4}}
e^{2C_1} \,\big(\eta_{\mu \nu} dx^\mu dx^\nu \big)    \Big) \,.} }
One can regularize the radial metric by setting $t \sim 
1 + \chi^{8/5}$, in which case the radial and $S^2$ part of
the metric become $d \chi^2 + \chi^2 d\Omega_2^2 $.
Thus the $2$-spheres are collapsing in a natural manner.
The metric in the $\theta$ direction is blowing up,  which
is reminiscent of approaching the ring singularity
of a rotating Kerr-Newman black-hole.  The scale factor
in front of the $D3$-branes is now vanishing, which suggests
that one can now access the far infra-red limit.

The foregoing limits of the dilaton and metric only
depend in a rather mild way upon $\lambda$ for $\lambda <1$,
and indeed are structurally identical to configurations with 
$\lambda=0$.  This means that the physical interpretation of
the ring singularity should be the same for all $\lambda <1$,
and suggests that if $\lambda <1$ then the gaugino vev
is becoming irrelevant to the infra-red structure, which
is then dominated by the flow in the mass, $m$.  In this
limit, the additional $U(1)$ symmetry is restored, and the
ring singularity is a duality averaged family of $5$-branes.
 
It is natural to wonder if the $5$-brane identification becomes
clearer for $\lambda =1$ since the ``ring of $5$-branes'' collapses
into two singularities at $\theta = \pm {\pi \over 4}$.   Indeed,
at these points one has $\tau =\pm 1$, which are not only rational,
but are consistent with  $(1, \pm 1)$ $5$-branes.   As we 
have already noted, the five-dimensional flows considered here are
self $S$-dual, and so finding such a pair of branes is the simplest
possible solution we could have found.  This solution no longer
has the ``unphysical'' $U(1)$ symmetry, and therefore makes sense
at finite $N$, and is also a good candidate for a string
background.  The asymptotic analysis of the metric in \vsingasympmet\ 
does not, however, appear to be  consistent with the $5$-brane 
interpretation.  In particular, the $\IR \IP^3$ remains round rather
than collapsing to an $S^2$.

Thus, if one approaches the core of the supergravity solution 
from a generic  direction one sees a $7$-brane, and the 
scale of the $D3$-brane world goes to a finite value.  If one 
approaches the core from a direction that is consistent
with having an infra-red vacuum in the field theory, then
one encounters some lower dimensional ``branes,'' and the scale 
in front of the  $D3$-branes can now run to the far infra-red.
If gaugino vev is too small then one finds a duality averaged  
ring of  $5$-branes in the core.  If the gaugino vev
is tuned to its maximum possible, and indeed critical, value then
the ``duality'' symmetry is not restored in the infra-red, and
core contains two discrete singularities with dilaton/axion $(p,q)$ 
charges of $(1,\pm 1)$.   The structure of the metric in this limit
appears rather different from that of $5$-branes wrapping an $S^2$.

\newsec{The $SU(3)$ invariant flow}

The complexity of the metric \intmetr\ makes it extremly difficult to
study the full lift of the GPPZ-flow to ten dimensions. We can,
however, consider a further truncation of this flow to a $SU(3)$
invariant subspace of the scalar manifold \scalmanQCD, obtained by
considering the flow of the $\sigma$ field only, \ie\ with $m$ set to
zero.  This yields an $\cN=1$ flow with the superpotential, cf.\
\GPPZWpot,
\eqn\sigW{
W(\sigma)\eql -{3\over 2} \cosh^2\sigma\,,
} 
and an  explicit solution given by \GPPZ\ (cf. \flowsoln)
\eqn\sigsolofflow{
\sigma(r) \eql {1\over 2}\log\,
        \left({1+\lambda t^3\over 1-\lambda t^3}\right)\,,
\qquad 
A(r) \eql {1\over 6} \log\left( t^{-6}-\lambda^2\right)
        +  \log({\ell\over L}) \,,
}
where
\eqn\sigpar{
t\eql e^{-r/L}\,,\qquad \lambda\eql e^C\,,
}
and where $C$ and $\ell$ are integration constants. 

One should note that the  $SU(3)$ invariant scalar submanifold  of 
this flow is the same as the submanifold parametrized by the $\chi$
field of the non-supersymmetric $SU(3)$ flow in
\refs{\GPPZold,\DistZam} and also the $\chi$ field of the LS-flow in
section 3. However,  the latter involves a different superpotential
and the fields, $\chi$ and $\rho$, form a coupled system in which 
a truncation to the $\chi$ field
alone is inconsistent. Indeed, if we set $\rho=1$ in \fieldeqs, the
only solution is $\chi=0$. Nevertheless, we may still use those
results from the lift of the LS-flow that do not depend explicitly on
the flow equations. 

The potential \PandW\ is, of course the same for the superpotentials
\Wreduced, with $\rho=1$, and \GPPZWpot, with $m=0$, when we identify
$\chi=\sigma$. There is a critical point of the potential at
$\sigma\equiv\chi={1\over 2}\log(2-\sqrt{3})$ \GRW, which corresponds
to the compactification of the chiral IIB supergravity for which the
internal manifold is a $U(1)$ bundle over $\IC \IP_2$ \LJR.\foot{See, 
\PopeCPtwo\ for a recent discussion of $U(1)$ bundles over $\IC\IP_n$'s.} 
The present flow turns out to be a simple deformation of that
solution.

This is rather easy to see if we work with the metric
\alchmetr. Consider the complex coordinates \complcoor\ and set 
\eqn\tosptwo{
u^i\eql u^3 \zeta^i \,,\quad i=1,2\,,\qquad
u^3=(1+\zeta^1\bar\zeta_1+\zeta^2\bar\zeta_2)^{-1/2} e^{i\phi}\,, }
where $\zeta^i$, $i=1,2$ are the standard complex coordinates on
$\IC\IP_2$ and $\phi$ is the coordinate along the $U(1)$ fiber of the
projection $S^5\rightarrow \IC\IP_2$. Convenient real coordinates are,
see e.g.\
\refs{\WarPope},
\eqn\coorcptwo{
\left(\matrix{\zeta^1\cr\zeta^2\cr}\right)
\eql
\tan\theta\,g(\al_1,\al_2,\al_3)
\left(\matrix{1\cr 0\cr}\right)\,,
} 
where, as usual, $\alpha_i$ are the $SU(2)$ Euler angles. The
ten-dimensional metric \newwarpmetr\ can now be recast into the 
following form:
\eqn\sigmetric{
ds_{10}^2\eql \cosh\sigma \left (e^{2A} dx_\mu
dx^\mu-dr^2\right) - L^2 \left(\cosh\sigma (d\phi-A_{\rm
FS})^2 + {1\over \cosh\sigma} ds_{\rm FS}^2\right)\,,
}
where $ds^2_{\rm FS}$ is the 
Fubini-Study metric on $\IC\IP_2$,
\eqn\sigintmet{ds^2_{{\rm FS}}  \eql
d\theta^2+{1\over 4}\sin^2\theta\left((\sigma_1)^2+(\sigma_2)^2
+\cos^2\theta\,(\sigma_3)^2\right)\,, }
and
\eqn\sigpotfs{
A_{{\rm FS}}\eql
{1\over 2}\sin^2\theta\,\sigma_3\,, }
is the $U(1)$ potential. We choose the 10-beins $e^M$,
$M=1,\ldots,10$, as follows
\eqn\tenbeins{
e^{\mu+1}\propto dx^\mu\,,\qquad
e^5\propto dr\,,\qquad e^6\propto d\theta\,,\qquad
e^{6+i}\propto\sigma_i\,,\qquad e^{10}\propto d\phi +\ldots\,. 
}

Recall that for the compactification in \LJR, the antisymmetric tensor
field is simply given by $G_{(3)}\propto
du^1\wedge du^2\wedge du^3$ with the potential
\eqn\potljr{
A_{\rm R}\eql {1\over 12} e^{3 i\phi}\,\sin\theta
\,\left(2i\, d\theta\wedge (\sigma_1+\sigma_2)+
        {1\over
        2}\sin(2\theta)(\sigma_1+i\sigma_2)\wedge\sigma_3\right)\,.
}
It has also been argued in \LJR\ (see, also \WarPope) that the $SU(3)$
symmetry essentially determines this potential up to an overall
scale. Thus, rather than starting with the result of section 3,
which would require passing to the other spherical coordinates, we
simply consider the following Ansatz:
\eqn\sigasanz{
G_{(3)}\eql dA_{(2)}\,, \qquad 
A_{(2)}\eql f_{(3)}\, A_{R}\,.
}
Similarly, we take
\eqn\sigfanz{
F_{(5)}\eql {\cal F}+*{\cal F}\,,\qquad
{\cal F}\eql dx^0 \wedge dx^1\wedge dx^2\wedge dx^3\wedge df_{(5)}\,.
}
Finally, the same calculation as in section 3 implies that the
dilaton/axion field vanishes and its field equation is satisfied
because of the chiral factor $\sigma_1+i\sigma_2$ in $G_{(3)}$.

To determine the two unknown functions $f_{(3)}(r)$ and $f_{(5)}(r)$,
we start with Einstein equations. The Ricci tensor is diagonal
\def\pmm{\phantom{-}}
\eqn\sigricci{\eqalign{
R_{MN}\eql f_1 \,{\rm diag}&
\left(\matrix{\pmm 
1, & -1, & -1, & -1, & -1, & \pmm 1, & \pmm 1, & \pmm 1, & \pmm 1, & \pmm
1 \cr}\right)\cr
+f_2 \,{\rm diag}&
\left(\matrix{\pmm 
0, & \pmm 0, & \pmm 0, & \pmm 0, & - 1, & \pmm 0, & \pmm 0, 
& \pmm 0, & \pmm 0, & \pmm
0 \cr}\right)\,,
\cr}
}
where
\eqn\ricfs{
f_1\eql {1\over 2L^2} \cosh(\sigma)(7+\cosh(2\sigma))\,,\qquad 
f_2\eql -{18\over L^2} \sinh\sigma\,\tanh\sigma\,,
}
as are the energy momentum tensors,
\eqn\sugemtha{\eqalign{
T^{(3)}_{11}\eql -T^{(3)}_{22}\eql \ldots \eql -T^{(3)}_{44}\eql  
T^{(3)}_{66}\eql
& \ldots \eql T^{(3)}_{99}\cr & \eql {\cosh\sigma\over 18 L^6}
(9f_{(3)}^2+L^2 (f'_{(3)})^2)\,,\cr}
}
\eqn\sugemthb{\eqalign{
T^{(3)}_{55}&\eql {\cosh\sigma\over 6 L^6}(-3f_{(3)}^2+L^2 (f'_{(3)})^2)\,,
\cr 
T^{(3)}_{10\,10}& \eql {\cosh\sigma\over 18 L^6}(27f_{(3)}^2-L^2 
(f'_{(3)})^2)\,,
\cr}}
and
\eqn\sigemfv{
T^{(5)}_{11}\eql -T^{(5)}_{22}\eql \ldots -T^{(5)}_{55}\eql T^{(5)}_{66}\eql
\ldots\eql T^{(5)}_{10\,10}\eql {4 e^{-4A} \,(f_{(5)}')^2
\over \cosh^5\sigma}\,.
}
where the $'$ denotes the derivative with respect to the flow
coordinate $r$.

Clearly, we should have $T^{(3)}_{66}=\ldots=T^{(3)}_{10\,10}$, which
yields 
\eqn\sigftheq{
f_{(3)}'\eql \pm {3\over L}\,f_{(3)}\,.
}
with the boundary conditions, $f_{(3)}(\infty)=f'_{(3)}(\infty)=0$.
The solution is $f_3(r)\eql C_3\,e^{-3r/L}$. 
Substituting this back into the Einstein equations we get, 
{\it a priori} three equations for $f_{(5)}'$
and  the integration constant $C_3$, but it turns out that they are
solved by
\eqn\sigffeq{
(f_{(5)}')^2\eql  {\ell^8\over L^{10}}\,
{e^{12r/L}\left(2 e^{6r/L}-3 e^{2C}\right)^2\over 4\,
\left( e^{6r/L}- e^{2C}\right)^{8/3}}\,,
}
and $C_3\eql 3\, L^2\,e^C$. 

Next we use the Maxwell equations which determine the sign of
$f_{(5)}'$. We also verify that the required Bianchi identities are
satisfied. Finally, integrating \sigffeq\ and re-expressing the result
as a function of $\sigma$ we obtain the following solution for the
antisymmetric tensor fields:
\eqn\sigassol{
f_{(3)}\eql 3L^2\tanh\sigma\,,\qquad
f_{(5)}\eql {\ell^4\over 4L^4} \lambda^{4/3} 
\cosh^{2/3}\sigma\coth^{4/3}\sigma\,.
}

We conclude with some comments about the formal properties of this
solution, in particular of the metric \sigmetric. We can recast it
in the form
\eqn\tenmet{
ds_{10}^2 \eql {1\over \sqrt{F}}\,( dx_\mu dx^\mu )
        -\sqrt{F}\,ds_6^2\,,
}
where the function $F$ is the analogue of the harmonic functions in
the ``brane-type'' solutions, and consider the metric
$ds_6^2$ on the six-dimensional manifold comprised of the flow
coordinate, $r$, and the internal manifold. It is easy to check that
\eqn\sixmet{
ds_6^2\eql a(\rho)^2 (dy)^2 + b(\rho)^2 (yJdy)^2\,,
}
where $y^i$ are unrestricted cartesian coordinates in $R^6$ and  
$\rho^2=y\cdot y$ is the radial variable related to the original flow
by
\eqn\therhosig{
\rho\eql \rho_0 \left(\coth{\sigma\over 2}\right)^{1/3}\,.
} The relation to the previous coordinates on the sphere is $y^I=\rho
x^I$. Setting $\rho_0=\lambda=1$, the functions $a(\rho)$ and 
$b(\rho)$ are given by
\eqn\andbofrh{
a(\rho)\eql {\ell\over 2^{1/3}}\,\left(1-{1\over \rho^6}\right)^{1/3}\,,
\qquad
b(\rho)\eql 2^{2/3} \ell\, {1\over (\rho^6-1)^{2/3}}\,,
}
and 
\eqn\theexpforf{
F\eql 2^{4/3} {L^4\over \ell^4} {\rho^4 (\rho^6-1)^{2/3}\over
(\rho^6+1)^2}\,. }
As expected, we find that $F$ is not harmonic with
respect to the six-dimensional metric \sixmet\ nor is the latter a
flat metric. However, $ds_6^2$ turns out to be Ricci flat -- a fact
that certainly should have some significance.

\newsec{Conclusions}

We now have several non-trivial lifts of five-dimensional supergravity
solutions to their ten-dimensional counterparts.  As was evident in
\FGPWb, and in the ``super-QCD'' flow presented here, it is essential
to work with the ten-dimensional solutions if one is to understand
properly the infra-red asymptotics of the supergravity descriptions
of these flows.  The five-dimensional solutions simply do not suffice.

An integral ingredient in understanding how to construct the lifts to
ten dimensions is the relationship between the ten-dimensional dilaton
and its five-dimensional counterpart.  As was remarked in \PWntwo, the
expression \DilAnsatz, in principle, provides us with an analytic
relation between the running gauge coupling, the $\cN=4$ coupling, the 
scale of the theory, and the running of the fermion and boson masses.  
In practice, the detailed interpretation of \DilAnsatz, and its
connection with an NSVZ beta function, is more vexatious.  The problem
is the precise relationship of supergravity and field theory
quantities, for example, the field theory scale and the supergravity
radius, or the invariants of the Higgs vevs, and the angular behaviour
of the supergravity solution.  There is also the possibility
of operator mixing, as we saw in the LS-flow.
In addition to this, it should also be
remembered that the supergravity solution is a strong coupling result,
and so it may not actually be possible to track the details all the
way to weak coupling results like the NSVZ beta-function.  Thus the
supergravity description exhibits all the proper structure, and
general behaviour, but detailed connections with the weak coupling
results remain elusive.

This raises the further question as to the extent that one should
expect to be able to probe the infra-red limit using the supergravity
solution.  The answer to this question seems to depend upon the
example.  For the LS flow the solution runs all the way to the new
critical point, and approaches a conformal theory.  Thus the
supergravity solution can ``integrate out'' the massive chiral
multiplet and access the region of the field theory at mass scales far
below the mass of the chiral multiplet.  For ``Flows to Hades'' the
supergravity approximation will break down near the singularity and so
from the naive, five-dimensional perspective one would expect that the
supergravity approximation will fail at some scale short of the
infra-red.  As was seen in
\PWntwo, and in most of the solutions here, the ten-dimensional 
solution can resolve structure in the singularity and sometimes allow
us to interpret the phase.

In this paper we saw how the ten-dimensional solution can also throw
up a new infra-red obstacle: the oxidation of the $D3$-branes into
$5$-branes and $7$-branes.  We saw in section 7 that for the
``super-QCD'' flow in which all the chiral multiplets are given the
same mass, $m_0$, the $D3$-brane throat generically ``rounds out'' 
into a $7$-brane at a radial coordinate value of $r \sim m_0 L$.  
However, for special directions on the $S^5$, the flow approaches a 
singularity that may be interpreted as a ring distribution of 
$5$-branes.   This meshes well with expectations from field theory in
that there is only a ground state in the infra-red if the vevs of the
complex scalar fields, $\Phi_j$, are real.   If this condition is
not met, then the flow runs into a ``brick wall,''  and the 
scale in front of the $D3$-brane part of the $7$-brane metric goes 
to a finite limit:   The infra-red limit in which the chiral multiplets
decouple is inaccessible.  On the other hand, if the vevs of the
the $\Phi_j$ are indeed real, then the flow runs to the ring of 
$5$-branes, and as has been argued in  \refs{\Myers,\PolStr}, 
the field theory  superpotential naturally leads to such 
dielectric $5$-branes.   Our results show some new elements
of this $5$-brane story:  First, if the gaugino
condensate is too small, we find the $5$-branes 
smeared out into  a ring.  This is because of a restoration
of a $U(1)$ duality symmetry in the infra-red, and the ring is a
``duality'' smeared family of $5$-branes.  If the gaugino
condensate runs with its critical initial value
({\it i.e.} maximum possible physical value) the flow does not 
``round out'' into the $7$-brane solution,
but  limits to some form of $(1,1)$ and $(1,-1)$ ``branes.''

We are thus brought back to the issue of how to get the ``correct''
flow in that it properly describes the phases of $\cN=1$
QCD.  Perhaps the most compelling features of the the dielectric
$5$-brane story  of \refs{\Myers,\PolStr} is that
it very naturally distinguishes between electric and magnetic
confinement in terms of the kinds of strings that can end upon
different species of $5$-brane.  In the five-dimensional
supergravity theory, this behaviour is still only visible as the result 
of some fine tuning.  This was apparent in \GPPZ\ where it was 
argued that Wilson loops exhibited confinement and 't Hooft loops 
exhibited screening. In the light of \discsymm\ we see that if there 
is indeed such behaviour for some Wilson loops, then Wilson loops 
that approach the core of the
solution from duality flipped ($u \leftrightarrow v$) direction will
exhibit the dual behaviour.  As a result, a real physical Wilson loop
will always be screened as it is lowered into the core of the
solution: if it approaches from the ``confining'' direction, it will
always be energetically favorable to change its orientation slightly,
and thereby screen the quarks.  In short, confining behaviour in the
solution of \GPPZ\ must be an artefact of fine tuning the direction of
approach, much like the confining behaviour found in \JMNW.  This very
general argument, based on \discsymm, shows that the ``super-QCD''
flow of \GPPZ\ cannot result in purely an $NS5$ or $D5$ brane in the
core.  Indeed the best we could do is find a $(1,1)$ brane paired
with a $(1,-1)$ brane.

{}From the detailed analysis of the vev  of the gaugino condensate
we have learnt that the structure of the infra-red limit is a
discontinuous function of the initial conditions of vevs.  
In the ten-dimensional solution this means that the infra-red 
physics will depend upon precisely what, and how, normalizable
modes are running.   Physically, given an infra-red vacuum, one
expects an exactly fixed relationships between the mass of the 
chiral multiplet and the vevs of various operators.
For the flows considered here, the natural choice is to take $\lambda
=1$: Our computations suggest that such flows appear to run to 
a solution that makes sense for finite $N$.  Therefore, in
seeking out the IR limit of the field theory it is tempting to take
the Holmesian approach of eliminating the impossible, and concluding
that whatever remains, however improbable, must be the truth: Namely
that the physical flow to the $\cN=1$ theory in the far infra-red must
be the one with $\lambda =1$. Indeed, a similar conclusion was reached
in \SGnaked, but for rather different reasons.
What we are missing is the possibility of running other
non-trivial vevs that, in ten-dimensions, correspond to
higher supergravity modes.   It is presumably the running of these
normalizable modes that makes the difference between a ring of
duality averaged $5$-branes, a pair of $(1,\pm 1)$ branes, and
a pure $D5$-brane, or pure $NS5$-brane.

\bigskip
\leftline{\bf Acknowledgements}

We would like to thank S.~Gubser, P.~Mayr, J.~Minahan, J.~Polchinski 
and A.~Zaffaroni for helpful conversations. We would particularly
like to thank E.~Witten for his encouragement and helpful comments during 
this project.  This work was supported
in part by funds provided by the DOE under grant number
DE-FG03-84ER-40168.

\appendix{A}{Explicit $G_{2(2)} $ Matrices}

\noblackbox

\leftline{\bf The $G_{2(2)}$ Matrix}

The following matrices generate $G_{2(2)}$ in its seven-dimensional
representation.  
 
$$
J_1~=~ \left(\matrix{ 0 & {\coeff{1}{2}} & 0 & 0 & 0 & 0 & 0 \cr
         -{\coeff{1}{2}} & 0 & 0 & 0 & 0 & 0 & 0 \cr 
        0 & 0 & 0 & -{\coeff{1}{2}} & 0 & 0 & 0 \cr 
        0 & 0 & {\coeff{1}{2}} & 0 & 0 & 0 & 0 \cr 
        0 & 0 & 0 & 0 & 0 & 0 & 0 \cr 
        0 & 0 & 0 & 0 & 0 & 0 & 0 \cr 
        0 & 0 & 0 & 0 & 0 & 0 & 0 \cr  } \right) \ ,
\qquad
J_2~=~ \left(\matrix{ 0 & 0 & {\coeff{1}{2}} & 0 & 0 & 0 & 0 \cr 
      0 & 0 & 0 & {\coeff{1}{2}} & 0 & 0 & 0 \cr
         {\coeff{1}{2}} & 0 & 0 & 0 & 0 & 0 & 0 \cr 
      0 & {\coeff{1}{2}} & 0 & 0 & 0 & 0 & 0 \cr 
      0 & 0 & 0 & 0 & 0 & 0 & 0 \cr 0 & 0 & 0 & 0 & 0 & 0 & 0 \cr 
      0 & 0 & 0 & 0 & 0 & 0 & 0 \cr  } \right)\,,
$$

$$
J_3~=~ \left(\matrix{ 0 & 0 & 0 & {\coeff{1}{2}} & 0 & 0 & 0 \cr 
      0 & 0 & -{\coeff{1}{2}} & 0 & 0 & 0 & 0 \cr 
      0 & -{\coeff{1}{2}} & 0 & 0 & 0 & 0 & 0 \cr
         {\coeff{1}{2}} & 0 & 0 & 0 & 0 & 0 & 0 \cr 
      0 & 0 & 0 & 0 & 0 & 0 & 0 \cr 0 & 0 & 0 & 0 & 0 & 0 & 0 \cr 
      0 & 0 & 0 & 0 & 0 & 0 & 0 \cr  } \right)\,,
$$

$$
K_1~=~ \left(\matrix{ 0 & {\coeff{1}{2}} & 0 & 0 & 0 & 0 & 0 \cr
         -{\coeff{1}{2}} & 0 & 0 & 0 & 0 & 0 & 0 \cr 
        0 & 0 & 0 & {\coeff{1}{2}} & 0 & 0 & 0 \cr 
        0 & 0 & -{\coeff{1}{2}} & 0 & 0 & 0 & 0 \cr 
        0 & 0 & 0 & 0 & 0 & 1 & 0 \cr 
        0 & 0 & 0 & 0 & -1 & 0 & 0 \cr 
        0 & 0 & 0 & 0 & 0 & 0 & 0 \cr  } \right) \ ,
\qquad
K_2~=~ \left(\matrix{ 0 & 0 & -{\coeff{1}{2}} & 0 & 0 & 0 & 0 \cr 
        0 & 0 & 0 & {\coeff{1}{2}} & 0 & 0 & 0 \cr
           -{\coeff{1}{2}} & 0 & 0 & 0 & 0 & 0 & 0 \cr 
        0 & {\coeff{1}{2}} & 0 & 0 & 0 & 0 & 0 \cr 
        0 & 0 & 0 & 0 & 0 & 0 & 1 \cr 
        0 & 0 & 0 & 0 & 0 & 0 & 0 \cr 
        0 & 0 & 0 & 0 & 1 & 0 & 0 \cr  } \right)\,,
$$

$$
K_3~=~ \left(\matrix{ 0 & 0 & 0 & {\coeff{1}{2}} & 0 & 0 & 0 \cr 
      0 & 0 & {\coeff{1}{2}} & 0 & 0 & 0 & 0 \cr 
      0 & {\coeff{1}{2}} & 0 & 0 & 0 & 0 & 0 \cr
         {\coeff{1}{2}} & 0 & 0 & 0 & 0 & 0 & 0 \cr 
      0 & 0 & 0 & 0 & 0 & 0 & 0 \cr 
      0 & 0 & 0 & 0 & 0 & 0 & -1 \cr 
      0 & 0 & 0 & 0 & 0 & -1 & 0 \cr  } \right)\,,
$$

$$
X_1~=~ \left(\matrix{ 0 & 0 & 0 & 0 & 0 & 0 & 0 \cr 
      0 & 0 & 0 & 0 & 0 & 0 & 0 \cr 0 & 0 & 0 & 0 & 1 & 0 & 0 \cr 
      0 & 0 & 0 & 0 & 0 & -1 & 0 \cr 
      0 & 0 & 1 & 0 & 0 & 0 & 0 \cr 
      0 & 0 & 0 & -1 & 0 & 0 & 0 \cr 
      0 & 0 & 0 & 0 & 0 & 0 & 0 \cr  } \right) \ ,
\qquad
X_2~=~ \left(\matrix{ 0 & 0 & 0 & 0 & 0 & 0 & 0 \cr 
      0 & 0 & 0 & 0 & 0 & 0 & 0 \cr 0 & 0 & 0 & 0 & 0 & 1 & 0 \cr 
      0 & 0 & 0 & 0 & 1 & 0 & 0 \cr 0 & 0 & 0 & 1 & 0 & 0 & 0 \cr 
      0 & 0 & 1 & 0 & 0 & 0 & 0 \cr 
      0 & 0 & 0 & 0 & 0 & 0 & 0 \cr  } \right)\,,
$$

$$
X_3~=~ \left(\matrix{ 0 & 0 & 0 & 0 & -1 & 0 & 0 \cr 
      0 & 0 & 0 & 0 & 0 & 1 & 0 \cr 0 & 0 & 0 & 0 & 0 & 0 & 0 \cr 
      0 & 0 & 0 & 0 & 0 & 0 & 0 \cr 1 & 0 & 0 & 0 & 0 & 0 & 0 \cr 
      0 & -1 & 0 & 0 & 0 & 0 & 0 \cr 
      0 & 0 & 0 & 0 & 0 & 0 & 0 \cr  } \right)\ ,
\qquad
X_4~=~ \left(\matrix{ 0 & 0 & 0 & 0 & 0 & 1 & 0 \cr 
        0 & 0 & 0 & 0 & 1 & 0 & 0 \cr 
        0 & 0 & 0 & 0 & 0 & 0 & 0 \cr 
        0 & 0 & 0 & 0 & 0 & 0 & 0 \cr 
        0 & -1 & 0 & 0 & 0 & 0 & 0 \cr -1 & 0 & 0 & 0 & 0 & 0 & 0 
          \cr 0 & 0 & 0 & 0 & 0 & 0 & 0 \cr  } \right)\,,
$$

$$
X_5~=~ \left(\matrix{ 0 & 0 & 0 & 0 & 0 & 0 & 1 \cr 
      0 & 0 & 0 & 0 & 0 & 0 & 0 \cr 0 & 0 & 0 & 0 & 1 & 0 & 0 \cr 
      0 & 0 & 0 & 0 & 0 & 0 & 0 \cr 0 & 0 & 1 & 0 & 0 & 0 & 0 \cr 
      0 & 0 & 0 & 0 & 0 & 0 & 0 \cr 
      1 & 0 & 0 & 0 & 0 & 0 & 0 \cr  } \right) \ ,
\qquad
X_6~=~ \left(\matrix{ 0 & 0 & 0 & 0 & 0 & 0 & 0 \cr 
      0 & 0 & 0 & 0 & 0 & 0 & 1 \cr 0 & 0 & 0 & 0 & 0 & 1 & 0 \cr 
      0 & 0 & 0 & 0 & 0 & 0 & 0 \cr 0 & 0 & 0 & 0 & 0 & 0 & 0 \cr 
      0 & 0 & 1 & 0 & 0 & 0 & 0 \cr 
      0 & 1 & 0 & 0 & 0 & 0 & 0 \cr  } \right)\,,
$$

$$
X_7~=~ \left(\matrix{ 0 & 0 & 0 & 0 & 0 & 0 & 0 \cr 
      0 & 0 & 0 & 0 & 0 & 1 & 0 \cr 0 & 0 & 0 & 0 & 0 & 0 & 1 \cr 
      0 & 0 & 0 & 0 & 0 & 0 & 0 \cr 0 & 0 & 0 & 0 & 0 & 0 & 0 \cr 
      0 & -1 & 0 & 0 & 0 & 0 & 0 \cr 
      0 & 0 & -1 & 0 & 0 & 0 & 0 \cr  }  \right)\ ,
\qquad
X_8~=~ \left(\matrix{ 0 & 0 & 0 & 0 & 0 & 0 & 0 \cr 
      0 & 0 & 0 & 0 & -1 & 0 & 0 \cr 
      0 & 0 & 0 & 0 & 0 & 0 & 0 \cr 0 & 0 & 0 & 0 & 0 & 0 & 1 \cr 
      0 & 1 & 0 & 0 & 0 & 0 & 0 \cr 0 & 0 & 0 & 0 & 0 & 0 & 0 \cr 
      0 & 0 & 0 & -1 & 0 & 0 & 0 \cr  } \right)\,.
$$
 
The subgroups $SL(2,\IR)_{5d}$
and $SL(2,\IR)_X$ are generated by $J_i$ and $K_i$ respectively.
The compact generators are $J_1, K_1, X_3,X_4,X_7$ and $X_8$,
while $J_1, J_2, J_3$, $K_1$ generate the invariances of the
potential.  The full scalar manifold can be parametrized by matrices
of the form:
\eqn\Matparam{\eqalign{M \eql \exp( & a_1 X_1 + a_2 X_2 + 
a_5 (X_5 - {\coeff{1}{2}} X_1)  + a_6 (X_6 - \coeff{1}{2} X_2) \cr & + 
a_7 J_2 + a_8 J_3 - a_3 K_2 + a_4 K_3) \ . }}

Now observe that:
$$
\eqalign{ \big[\, K_1,\, (X_5 - \coeff{1}{2} X_1)\, \big] \eql & 
- (X_6 - \coeff{1}{2} X_2)\,, \cr 
 \big[\, K_1,\, (X_6 - \coeff{1}{2} X_2)\, \big] \eql & 
 (X_5 - \coeff{1}{2} X_1) \,.}
$$
We can therefore use the $K_1$ invariance to take $a_6 = 0$.  
We can then use the $SL(2,\IR)_{5d}$ to set $a_7 = -a_3$
and $a_8 = a_4$.  Doing this we get a five-parameter family
of matrices, with an unused $J_1$ invariance.  Introduce the
following change of basis matrix:
$$
B~=~ \left(\matrix{ 0 & 0 & 1 & 0 & 1 & 0 & 0 \cr 
      0 & 0 & 0 & 1 & 0 & 1 & 0 \cr 1 & 0 & 0 & 0 & 0 & 0 & -1 \cr 
      0 & 0 & 1 & 0 & -1 & 0 & 0 \cr 0 & 0 & 0 & 1 & 0 & -1 & 0 \cr 
      1 & 0 & 0 & 0 & 0 & 0 & 1 \cr 
      0 & 1 & 0 & 0 & 0 & 0 & 0 \cr  }  \right)\ ,
$$
then
$$
B~M~B^{-1} ~=~ \left(\matrix{ P & 0 & 0  \cr  0 & -P & 0  
\cr 0 & 0 & 1 \cr}  \right)\,, \quad {\rm where} \quad
P ~=~ \left(\matrix{a_1 + \half a_5 & a_2 & a_3  \cr  a_2 & -a_1 + 
\half a_5 & a_4  \cr a_3 & a_4 & -a_5 \cr}  \right) \,. 
$$
This explicitly defines the embedding of the non-compact
part of $SL(3,\IR)$ into $G_{2(2)}$.  

Finally, let $H_1 = -(J_1 - K_1)$, $H_2 = (X_3 - X_7)$ and 
$H_3 = (X_4 + X_8)$.  Then these matrices define the $SO(3)$
subgroup of the foregoing $SL(3,\IR)$ into $G_{2(2)}$, and indeed,
$$
B~\exp\bigg(\sum_{j=1}^3 ~c_j ~H_j\bigg)~B^{-1} ~=~ 
\left(\matrix{ A & 0 & 0  \cr  0 & -A & 0  \cr 0 & 0 & 1 \cr}  \right)\,, 
\quad {\rm where} \quad A ~=~ \left( \matrix{0 & c_1 & c_2  
\cr  -c_1 & 0 & c_3  \cr -c_2 & -c_3 & 0 \cr}\right) \,.
$$

\listrefs 
\vfill
\eject
\end